\documentclass[oupdraft]{bio} 
\usepackage{url}

\newcommand{\I}{\mathrm{I}}

\newcommand{\Var}{\mathbb{V}\mathrm{ar}}
\newcommand{\TCF}{\mathrm{TCF}}

\newcommand{\asyD}{\stackrel{d}{\longrightarrow}}

\begin{document}

\title{Interval estimation in three-class ROC analysis: a fairly general approach based on the empirical likelihood}

\author{DUC-KHANH TO$^{\ast, 1,2}$, GIANFRANCO ADIMARI$^1$, MONICA CHIOGNA$^3$\\[4pt]
\textit{$^1$Department of Statistical Sciences, University of Padova, Via C. Battisti, 241; I-35121 Padova, Italy} \\
\textit{$^2$Department of Information and Engineering, University of Padova, Via Gradenigo, 6/b; 35131 Padova, Italy} \\
\textit{$^3$Department of Statistical Sciences ``Paolo Fortunati'', University of Bologna, Via Belle Arti, 41; 40126 Bologna, Italy}
\\[2pt]
{duckhanh.to@unipd.it}}

\markboth%
{D.-K. To, G. Adimari and M. Chiogna}
{Empirical likelihood-based interval estimation in three-class ROC analysis}

\maketitle

\footnotetext{To whom correspondence should be addressed.}

\begin{abstract}
{The empirical likelihood is a powerful nonparametric tool, that emulates its parametric counterpart -the parametric likelihood- preserving many of its large-sample properties. This article tackles the problem of assessing the discriminatory power of three-class diagnostic tests from an empirical likelihood perspective. In particular, we concentrate on interval estimation in a three-class ROC analysis, where a variety of inferential tasks could be of interest. We present novel theoretical results and tailored techniques studied to efficiently solve some of such tasks. Extensive simulation experiments are provided in a supporting role, with our novel proposals compared to existing competitors, when possible. It emerges that our new proposals are extremely flexible, being able to compete with contestants and being the most suited to accommodating flexible distributions for target populations. We illustrate the application of the novel proposals with a real data example. The article ends with a discussion and a presentation of some directions for future research.}
{Bootstrap; Diagnostic test; Nonparametric inference; ROC surface; VUS}
\end{abstract}

\section{Introduction}
\label{sec:into}
The construction of confidence intervals or regions for unknown parameters of interest is a classic problem of statistical inference, and, by the time, the approach based on the empirical likelihood (EL) (see \citealp{owen2001empirical}, as general reference) has shown its effectiveness and flexibility in addressing such kind of problem. 

The empirical likelihood is a nonparametric tool that allows obtaining pseudo-likelihoods in several contexts and, in particular, for parameters that are determined by estimating equations. By an emulation of its parametric counterpart, the empirical likelihood function is obtained by maximization of a nonparametric likelihood supported on the data, subject to some constraints. In most cases, the maximization problem is solved by using Lagrange multipliers. This leads to an explicit expression for (minus twice) the empirical log-likelihood ratio, for which a Wilks-type theorem is generally proved. Then, the empirical likelihood can be used, in a standard way, to obtain non-parametric confidence intervals or regions, which, by their nature, are range-respecting and have the shape determined only by data, free from any artificial constraint, such as that of symmetry. Moreover, EL regions are generally more accurate than traditional ones, based on the asymptotic normality of some estimator for the parameter of interest. Finally, in most complex situations empirical likelihood admits simplified versions, obtained by plugging-in appropriate estimates of parameters, often nuisance parameters. These versions benefit from the reduced computational burden but at the expense of the shape of the approximant distribution, which often changes from the standard $\chi^2$ distribution to a scaled $\chi^2$ distribution (see \citealp{hjort2009extending}, and \citealp{adimari2010note}). In the following, we will call one such version ``estimated empirical likelihood''.

Nowadays, EL methods have a wide range of applications in various research fields (see, \citealp{lazar2021review} and \citealp{liu2022review}, for recent reviews), including the receiver operating characteristic (ROC) analysis for a three-class problem, which is commonly used to evaluate the ability of a diagnostic test (or biomarker) to distinguish three ordinal classes (e.g., benign, stage 1, stage 2) of a disease \citep{nak:14}. 

Let $Y$ be a diagnostic test result often measured on a continuous scale, and let $Y_1, Y_2, Y_3$ be the test result for subjects in classes 1, 2, and 3, respectively. Without loss of generality, we assume that higher values of test results are associated with higher severity of the disease. This assumption can be formalized by the simple ordering of the means, i.e., $\mu_1 < \mu_2 < \mu_3$, here $\mu_j$ is the mean of $Y_j$ ($j = 1, 2, 3$). Given a pair of thresholds $(t_1, t_2)$, with $t_1 < t_2$, the true class fractions (TCFs) are defined as \citep{nak:04}
\begin{eqnarray}
\theta_1 \equiv \TCF_1(t_1) &=& \Pr(Y_1 \le t_1) = F_1(t_1) \nonumber \\
\theta_2 \equiv \TCF_2(t_1, t_2) &=& \Pr(t_1 < Y_2 \le t_2) = F_2(t_2) - F_2(t_1) \label{eq:tcfs} \\
\theta_3 \equiv \TCF_3(t_2) &=& \Pr(Y_3 > t_2) = 1 - F_3(t_2) \nonumber
\end{eqnarray}
where $F_1(\cdot)$, $F_2(\cdot)$ and $F_3(\cdot)$ are the cumulative distribution functions of $Y_1$, $Y_2$ and $Y_3$, respectively. Then, by plotting $(\TCF_1(t_1), \TCF_2(t_1, t_2), \TCF_3(t_2))$ in a unit cube over all possible values of thresholds $t_1 < t_2$, one obtain the ROC surface for the test $Y$ \citep{nak:04}. The volume under the ROC surface (VUS), defined as 
\begin{equation}
\gamma = \Pr(Y_1 < Y_2 < Y_3),
\end{equation}
is usually considered a summary measure of the diagnostic accuracy of the test. The values of VUS vary from $1/6$ to 1, ranging from the chance to perfect diagnostic tests \citep{nak:04}.

Several parametric, semi-parametric, or kernel-based approaches have been developed to estimate the ROC surface; we cite here, among others, papers by \citet{nak:04}, \citet{xiong2006measuring}, \citet{li2009nonparametric} and \citet{kang2013estimation}. The asymptotic normality of the proposed estimators can be used to construct confidence regions at a fixed point of the ROC surface, i.e.,
for a  tern of TCFs, $(\theta_1, \theta_2, \theta_3)$, at a given pair of thresholds $(t_1, t_2)$. To the best of our knowledge, no EL methods are available to attain this last goal.

As for the construction of confidence intervals for the VUS, parametric or nonparametric approaches have been developed; for instance, by \citet{xiong2006measuring,guangming2013nonparametric,nak:14}. Apart from the ``hybrid'' approach proposed by \citet{guangming2013nonparametric} which uses jackknife empirical likelihood \citep{jing2009jackknife}, only a ``proper'' (estimated) EL method has been proposed by \citet{wan2012empirical}, based on the so-called placement values.  The author defined an empirical likelihood pivot and proved the appropriateness of an approximant scaled $\chi^2$ distribution, with an unknown scale constant. Such constant results in a ratio of variances, and can be estimated through a ratio of functions of U-statistics.

Within the three-class ROC analysis framework, the problem of making inferences about a TCF, given the remaining two, is also of interest. Especially, for medical practitioners can be important to evaluate the accuracy of a diagnostic test to distinguish the second class of disease, i.e., to evaluate the probability, $\theta_2$, with which the test correctly classifies a subject at the second stage of disease (often called early stage), when are fixed the values for the true class fractions at first and third classes, $\theta_1$ and $\theta_3$, i.e., when the test simultaneously ensures certain values for the correct probabilities of classification for the first and third stages. 

In the last decade, methods have been developed to address the problem of building confidence intervals for $\theta_2$, given $\theta_1$ and $\theta_3$. \citet{dong2011parametric} proposed to use a generalized inference approach for interval estimation of $\theta_2$, under normality assumption or after Box-Cox transformation in non-normal cases. The authors also developed some nonparametric bootstrap-based approaches, referred to as BTP and BTII. \citet{dong2015confidence}, instead,  developed two estimated empirical likelihood-based methods, ELP and ELB. The authors defined an estimated profile empirical log-likelihood ratio for $\theta_2$ and proved that it approximately follows a scaled $\chi^2$ distribution, with an unknown scale constant which is still a ratio of variances. To estimate the unknown scale constant, the authors proposed, as alternatives, a method involving kernel density estimators, and a bootstrap-based procedure. A different profile empirical log-likelihood for $\theta_2$, and an adjusted version, have been proposed by \citet{rahman2022empirical}, referred to as PEL and AEL. They have the advantage of having the standard $\chi^2$ distribution as approximating distribution, but they are more complicated to compute than the competitors mentioned above. Finally, \citet{hai2022bayesian} proposed to obtain a confidence interval for $\theta_2$ (given $\theta_1$ and $\theta_3$) by using an estimated empirical likelihood pivot based on an estimated version of the so-called influence function of an estimator for $\theta_2$. The proposed pivot has a standard $\chi^2$ asymptotic distribution, but the estimation of the influence function involves kernel density estimation.

In summary, the above-mentioned empirical likelihood techniques have different genesis and solve the problems of building approximate confidence intervals for the TCF at the early stage of disease and for the VUS of a diagnostic test. No EL methods to construct confidence regions for the tern  $(\theta_1,\theta_2,\theta_3)$ on the ROC surface, at a fixed pair of thresholds $(t_1,t_2)$, are present in the literature. Moreover, to the best of our knowledge, there are no methods in the literature for constructing confidence regions for a pair of TCFs, the remaining other fixed. In practice, there may be situations in which, for instance, the decision to treat a patient because erroneously positive to the response of a diagnostic test could have high costs (personal or for the health system). In such situations, the researcher may want to require the test to have some fixed level for $\TCF_1$ (the first class typically considers the absence of disease or its harmless level) and study its ability to discriminate between the second and third classes of disease.

This paper aims to propose a unified EL approach that allows to solve the four types of problems considered so far. Our proposal provides EL techniques that are generally easier to use when compared to competing ones. Moreover, the results of an extensive simulation study indicate that our techniques are generally at least as accurate as others (not just those based on EL), and sometimes more accurate in terms of real coverage in finite samples.

The paper is organized as follows. In Section 2, the proposals are laid out in detail, supported by theoretical results. Section 3 presents extensive simulation studies to assess our proposals' performance in finite samples, and compare them with existing methods when possible. Section 4 illustrates our methods in evaluating the ability of some gene expressions to distinguish ductal carcinomas in situ from noncancerous and invasive breast cancers. Finally, a few concluding remarks can be found in Section 5, along with some considerations about various directions for future research.

\section{The proposal}
\label{sec:2}
\subsection{Three-sample empirical likelihood and confidence regions for triplets of TCFs}
\label{sec:2:1}
Let $\left\{y_{d1}, \ldots, y_{dn_d}\right\}$  be a random sample from $Y_d$, the test result for $n_d$ patients from the $d$-th class, $d = 1,2,3$. For a fixed $t$, let $\widehat{F}_d(t) = \frac{1}{n_d} \sum\limits_{i = 1}^{n_d} \I(y_{di} \le t)$, be the empirical distribution functions based on the sample $\left\{y_{d1}, \ldots, y_{dn_d}\right\}$, and $\widehat{P}(t_1, t_2) = \widehat{F}_2(t_2) - \widehat{F}_2(t_1)$ for fixed $t_1$ and $t_2$ ($t_1 < t_2$), where $\I(\cdot)$ is the indicator function. Let $\boldsymbol{p}_d = (p_{d1}, \ldots, p_{dn_d})$ denote a probability vector, with $p_{di} > 0$, parameterizing multinomial distributions on the data points $\left\{y_{d1}, \ldots, y_{dn_d}\right\}$, with $d = 1,2,3$. The empirical likelihood for $(\theta_1, \theta_2, \theta_3)$, at fixed $t_1<t_2$, can be defined as
\begin{equation}
L(\theta_1, \theta_2, \theta_3; t_1, t_2) = \sup_{\boldsymbol{p}_1, \boldsymbol{p}_2, \boldsymbol{p}_3} \prod_{i = 1}^{n_1}p_{1i}\prod_{j = 1}^{n_2}p_{2j} \prod_{k = 1}^{n_3}p_{3k}
\label{emp:like}
\end{equation}
subject to the following constraints: $p_{1i} > 0$, $p_{2j} > 0$, $p_{3k} > 0$, $\sum\limits_{i = 1}^{n_1}p_{1i} = 1$, $\sum\limits_{j = 1}^{n_2}p_{2j} = 1$, $\sum\limits_{k = 1}^{n_3}p_{3k} = 1$,
\[
\sum_{i = 1}^{n_1}p_{1i}\I(y_{1i} \le t_1) = \theta_1, \quad \sum_{j = 1}^{n_2}p_{2j}\I(t_1 < y_{2j} \le t_2) = \theta_2, \quad \sum_{k = 1}^{n_3}p_{3k}\I(y_{3k} \le t_2) = 1 - \theta_3.
\]
Under the constraints $\sum\limits_{i = 1}^{n_1}p_{1i} = 1$, $\sum\limits_{j = 1}^{n_2}p_{2j} = 1$ and $\sum\limits_{k = 1}^{n_3}p_{3k} = 1$, the empirical likelihood $L(\theta_1, \theta_2, \theta_3; t_1, t_2)$ in (\ref{emp:like}) will reach its maximum $n_1^{n_1} n_2^{n_2} n_3^{n_3}$, at $p_{1i} = 1/n_1$, $p_{2j} = 1/n_2$ and $p_{3k} = 1/n_3$ for all $i, j, k$. Thus, the empirical log-likelihood ratio for $(\theta_1, \theta_2, \theta_3)$ is defined as
\begin{equation}
\ell(\theta_1, \theta_2, \theta_3; t_1, t_2) = \sup_{\boldsymbol{p}_1, \boldsymbol{p}_2, \boldsymbol{p}_3} \left\{-2\sum_{i = 1}^{n_1}\log\left(n_{1i}p_{1i}\right) - 2\sum_{j = 1}^{n_2} \log\left(n_2p_{2j}\right) - 2\sum_{k = 1}^{n_3} \log\left(n_3p_{3k}\right)\right\},
\label{emp:like:ratio}
\end{equation}
subject to the constraints mentioned above.
In Section A of the Supplementary Materials, we show that, after some algebra,
\begin{eqnarray} 
\ell(\theta_1,\theta_2,\theta_3,;t_1, t_2) &=& 2n_1 \left\{\widehat{F}_1(t_1) \log \frac{\widehat{F}_1(t_1)}{\theta_1} + \left[1 - \widehat{F}_1(t_1)\right] \log\frac{1 - \widehat{F}_1(t_1)}{1 - \theta_1} \right\} \nonumber \\ 
&& + \: 2n_2 \left\{\widehat{P}(t_1, t_2) \log \frac{\widehat{P}(t_1, t_2)}{\theta_2} + \left[1 - \widehat{P}(t_1, t_2)\right] \log\frac{1 - \widehat{P}(t_1, t_2)}{1 - \theta_2} \right\} \nonumber \\
&& + \: 2n_3 \left\{\widehat{F}_3(t_2) \log \frac{\widehat{F}_3(t_2)}{1 - \theta_3} + \left[1 - \widehat{F}_3(t_2)\right]\log\frac{1 - \widehat{F}_3(t_2)}{\theta_3} \right\},
\label{emp:like:3C}
\end{eqnarray}
when
\[
\begin{cases}
t_1 \in \left[y_{1(1)}, y_{1(n_1)}\right), \\
t_1 \text{ or } t_2 \in \left[y_{2(1)}, y_{2(n_2)}\right), \\
t_2 \in \left[y_{3(1)}, y_{3(n_3)}\right),
\end{cases}
\]
otherwise $\ell(\theta_1,\theta_2,\theta_3;t_1,t_2) = +\infty$. It is worth noting that the empirical estimate $\widehat{P}(t_1, t_2)$ can be zero when there are no observations $y_{2j}$ (with $j = 1, 2, \ldots, n_2$) laying between $(t_1, t_2)$; in this case $\ell(\theta_1,\theta_2,\theta_3;t_1,t_2)$ is also not well-defined. This can happen when the sample size $n_2$ is very small, and to avoid this drawback, one could use in (\ref{emp:like:3C}) continuous versions of $\widehat{F}_d$, as suggested in \citet{adimari1998empirical}.

The next theorem establishes the asymptotic behavior of $\ell(\theta_1,\theta_2,\theta_3;t_1,t_2)$. For a fixed pair of thresholds $(t_{10}, t_{20})$ such that $t_{10} < t_{20}$, we denote as $\theta_{10} = F_1(t_{10})$, $\theta_{20} = F_2(t_{20}) - F_2(t_{10})$ and $\theta_{30} = 1 - F_3(t_{20})$ the true values for the parameters of interest. 
\begin{theorem}
\label{thrm:1}
If $\min\{n_1, n_2, n_3\} \to +\infty$, then we have
\begin{equation}
\ell(\theta_{10},\theta_{20},\theta_{30};t_{10},t_{20}) \asyD \chi^2_3,
\label{asy:ll:3C}
\end{equation}
where $\chi^2_3$ indicates the chi-squared distribution with 3 degrees of freedom.
\end{theorem}
The proof can be found in Section B of the Supplementary Materials. Based on the result in Theorem \ref{thrm:1}, we can construct a set 
\[
{\cal R}_{\alpha}=\left\{(\theta_1, \theta_2, \theta_3): \ell(\theta_1, \theta_2, \theta_3; t_{10}, t_{20}) \le \chi_{3, (1 - \alpha)}^2 \right\},
\]
where $\alpha \in (0,1)$ and $\chi_{3, (1 - \alpha)}^2$ is the $(1 - \alpha)$-th quantile of a $\chi_3^2$ distribution. ${\cal R}_{\alpha}$ is a (nonparametric) confidence region, with nominal coverage probability $1 - \alpha$, for a TCFs triplet $(\theta_{10}, \theta_{20}, \theta_{30})$, at a fixed pair of thresholds $(t_{10}, t_{20})$. This 3D confidence region can be easily produced  using the \texttt{contour3d()} function of the \texttt{R} \citep{R_team} package \texttt{misc3d} \citep{misc3d}. In what follows, we will denote such confidence regions as  ELQ3D.

{
It is worth noting that the nonparametric log-likelihood ratio in (\ref{emp:like:3C}), coincides with the log-likelihood ratio based on the (independent) binomial distributions of $\widehat{F}_1(t_1)$, $\widehat{P}_2(t_1, t_2)$ and $\widehat{F}_3(t_2)$. 
This provides a further justification for the result in Theorem 2.1 under the considered weak conditions, which do not even involve the continuity of $F_1$, $F_2$ and $F_3$.
}

\subsection{Confidence intervals for TCF$_2$}
\label{sec:2:2}
In some circumstances, one could have enough information to fix the values for $\theta_1$ and $\theta_3,$ or one may want to fix such quantities to desired values. Then, thresholds $t_1$ and $t_2$ could be estimated. Let $\widehat{t}_1 = \widehat{F}_1^{-1}(\theta_1)=\inf\{t:\widehat{F}_1(t)\ge\theta_{10}\}$ and $\widehat{t}_2 = \widehat{F}_3^{-1}(1 - \theta_3)$ be estimates of $t_1$ and $t_2$, respectively, for fixed $\theta_1$ and $\theta_3$. In this situation, by using the plug-in method, we have an estimated version of the empirical likelihood statistic $\ell(\theta_1, \theta_2, \theta_3; t_1, t_2)$ as follows:
\begin{eqnarray}
\ell_*(\theta_2) &=& \ell(\theta_1, \theta_2, \theta_3; \widehat{t}_1, \widehat{t}_2) \nonumber \\
&=& 2n_1 \left\{\widehat{F}_1(\widehat{t}_1) \log \frac{\widehat{F}_1(\widehat{t}_1)}{\theta_1} + \left[1 - \widehat{F}_1(\widehat{t}_1)\right] \log\frac{1 - \widehat{F}_1(\widehat{t}_1)}{1 - \theta_1} \right\} \nonumber \\ 
&& + \: 2n_2 \left\{\widehat{P}(\widehat{t}_1, \widehat{t}_2) \log \frac{\widehat{P}(\widehat{t}_1, \widehat{t}_2)}{\theta_2} + \left[1 - \widehat{P}(\widehat{t}_1, \widehat{t}_2)\right] \log\frac{1 - \widehat{P}(\widehat{t}_1, \widehat{t}_2)}{1 - \theta_2} \right\} \nonumber \\
&& + \: 2n_3 \left\{\widehat{F}_3(\widehat{t}_2) \log \frac{\widehat{F}_3(\widehat{t}_2)}{1 - \theta_3} + \left[1 - \widehat{F}_3(\widehat{t}_2)\right]\log\frac{1 - \widehat{F}_3(\widehat{t}_2)}{\theta_3} \right\},
\label{emp:like:2_2}
\end{eqnarray}
if $\widehat{t}_1$ or $\widehat{t}_2 \in \left[y_{2(1)}, y_{2(n_2)}\right)$, otherwise $\ell_*(\theta_2) = +\infty$. Substitution of $t_1$ and $t_2$ by their estimates impacts on the standard $\chi^2$ approximation, that no longer holds. The following theorem shows that $\ell_*(\theta_{20})$, for fixed $\theta_1=\theta_{10}$ and $\theta_3=\theta_{30}$, asymptotically follows a scaled $\chi^2$ distribution, under some smoothness conditions. Again, $\theta_{20}$ denotes the true value for the parameter of interest.

\begin{theorem}
\label{thrm:tcf2}
Assume that $F_1$, $F_2$ and $F_3$ have continuous density functions $f_1$, $f_2$ and $f_3$ such that $f_1(t_{10}) > 0$, $f_2(t_{10}) > 0$, $f_2(t_{20}) > 0$ and $f_3(t_{20}) > 0$. Let $t_{10} = F_1^{-1}(\theta_{10})$ and $t_{20}=1-F_3^{-1}(\theta_{30})$. If $\min\{n_1, n_2, n_3\} \to +\infty$ and the ratios $n_1/n_2$, $n_3/n_2$ have finite non-zero limits, then
\begin{eqnarray}
w\ell_*(\theta_{20}) = w\ell(\theta_{10}, \theta_{20}, \theta_{30}; \widehat{t}_{10}, \widehat{t}_{20}) \stackrel{d}{\longrightarrow} \chi^2_1,
\label{emp:like:2_3}
\end{eqnarray}
where $w > 0$ is a suitable finite parameter.
\end{theorem}

The proof can be found in Section C of the Supplementary Materials. To estimate $w$, one could use estimators for variances. However, we observe that, from (\ref{emp:like:2_3}), the quantity $w\mathrm{Med}(\ell_*(\theta_{20}))$ is asymptotically equal to $\mathrm{Med}(\chi_1^2)$ which is $(7/9)^3$; here $\mathrm{Med}(\cdot)$ stands for the median operator. Thus, we propose to estimate $w$ by $\widehat{w} = \dfrac{(7/9)^3}{\widehat{\mathrm{Med}}(\ell_*(\theta_{20}))}$, with $\widehat{\mathrm{Med}}(\ell_*(\theta_{20}))$ obtained by the following bootstrap procedure:
\begin{enumerate}
\item from observed data $y_{11},\ldots, y_{1n_1}$, $y_{21},\ldots, y_{2n_2}$ and $y_{31},\ldots, y_{3n_3}$, obtain the estimates $\widehat{t}_{10}$, $\widehat{t}_{20}$ and $\widehat{\theta}_{20} = \widehat{F}_2(\widehat{t}_{20}) - \widehat{F}_2(\widehat{t}_{10})$;
\item get $B$  bootstrap samples $\{y_1\}_{b}$, $\{y_2\}_{b}$ and $\{y_3\}_{b}$, for $b = 1, \ldots, B$, of sizes $n_1$, $n_2$ and $n_3$, respectively;
\item from the $b$-th pair of bootstrap samples, compute the estimates $\widehat{t}_{10b}$ and $\widehat{t}_{20b}$, and then, $\ell_{*b}(\widehat{\theta}_{20}) = \ell(\theta_{10}, \widehat{\theta}_{20}, \theta_{30}; \widehat{t}_{10b}, \widehat{t}_{20b})$;
\item get the estimates $\widehat{\mathrm{Med}}(\ell_*(\theta_{20}))$ as the sample medians from the values $\ell_{*b}(\widehat{\theta}_{20})$, $b = 1,\ldots, B$.
\end{enumerate}
In the above-given procedure, only those bootstrap samples are processed whose sample averages respect the fixed ordering of the population means ($\mu_1 < \mu_2 < \mu_3$).  To improve accuracy in the estimation of $w$, in step 4 the continuous version of $\widehat{F}_j$  \citep{adimari1998empirical} is employed when computing, from each bootstrap sample, $\ell_{*b}(\widehat{\theta}_{20})$. { The choice to use bootstrap to estimate the median of $\ell_{*}(\theta_{20})$ is dictated by the simplicity of this approach with respect, for instance, to the bootstrap calibration, and by the fact that resorting to the median avoids suitable {\it ad hoc} measures to treat situations in which $\ell_{*b}(\widehat{\theta}_{20})$ assumes not finite value.} 

A confidence interval for $\theta_{20}$, for fixed $\theta_1=\theta_{10}$ and $\theta_3=\theta_{30}$, with nominal coverage $1-\alpha$, is therefore obtained as
\[
{\cal R}^*_{2,\alpha} = \left\{\theta_2: \widehat{w} \ell_*(\theta_2) \le \chi_{1, (1 - \alpha)}^2 \right\},
\]
where $\alpha \in (0,1)$ and $\chi_{1, (1 - \alpha)}^2$ is the $(1 - \alpha)$-th quantile of a $\chi_1^2$ distribution. The interval ${\mathcal{R}}_{2,\alpha}^*$ will be denoted as ELQB.

\subsection{Confidence intervals for the VUS}
\label{sec:2:3}
Observe that, in (\ref{emp:like:3C}), each of the three addenda is an empirical likelihood pivot (with approximant distribution $\chi^2_1$) for inference about an unknown proportion or probability,
estimated by its empirical counterpart; in a sample of independent and identically distributed observations, such empirical counterpart, multiplied by the sample size, is the realization of a binomial random variable.

Let $\gamma$ be the unknown VUS  for a diagnostic test. As $\gamma$ is a probability, our idea is to use the quantity
\begin{equation}
\ell(\gamma) = 2n\left\{\widehat{\gamma} \log \frac{\widehat{\gamma}}{\gamma} + (1 - \widehat{\gamma}) \log \frac{1 - \widehat{\gamma}}{1 - \gamma}\right\},
\label{ell:vus}
\end{equation}
to obtain a pivot for interval estimation of the VUS. In (\ref{ell:vus}), 
$n = n_1 + n_2 + n_3$ and
\begin{equation}
\widehat{\gamma} = \widehat{\Pr}(Y_1 < Y_2 < Y_3) = \frac{1}{n_1 n_2 n_3}\sum_{i = 1}^{n_1} \sum_{j = 1}^{n_2} \sum_{k = 1}^{n_3} \I\left(y_{1i} < y_{2j} < y_{3k}\right),
\label{est:vus}
\end{equation}
is an estimate, given by an unbiased nonparametric estimator \citep{nak:04}. Assuming $\gamma \in (0,1)$,  $\ell(\gamma)$  in (\ref{ell:vus}) is well-defined if $\widehat{\gamma}$ is different from 1 (or 0). However, the estimate $\widehat{\gamma}$ is a three-sample U-statistic, and its presence in (\ref{ell:vus}) affects the asymptotic behaviour of the quantity itself. We prove the following theorem.

\begin{theorem}
\label{thrm:vus}
Let $\gamma_0$ be the true value of $\gamma \in (0,1)$.
When $\min\{n_1, n_2, n_3\} \to +\infty$ and $n_d/n \to \rho_d$, with $0 < \rho_d < 1$, \ $d = 1,2,3$, then
\begin{equation}
w\ell(\gamma_0) \stackrel{d}{\longrightarrow} \chi^2_1,
\label{emp:like:vus}
\end{equation}
where $w > 0$ is a suitable finite parameter.
\end{theorem}

The proof can be found in Section D of the Supplementary Materials. One could estimate the scale parameter $w$ as $\dfrac{\widehat{\gamma}(1 - \widehat{\gamma})}{n \, \widehat{\Var}(\widehat{\gamma})}$, where $\widehat{\Var}(\widehat{\gamma})$ is the estimated variance of $\widehat{\gamma}$ \citep{lehmann1998elements}. However, we follow the same idea as in Section \ref{sec:2:2}, and propose to estimate $w$ by $\widehat{w} = \dfrac{(7/9)^3}{\widehat{\mathrm{Med}}(\ell(\gamma_0))}$, with the estimate $\widehat{\mathrm{Med}}(\ell(\gamma_0))$ obtained by the following bootstrap procedure:
\begin{enumerate}
\item from the observed data $y_{11},\ldots, y_{1n_1}$, $y_{21},\ldots, y_{2n_2}$ and $y_{31},\ldots, y_{3n_3}$, obtain the estimate $\widehat{\gamma}$ by  (\ref{est:vus});
\item get $B$ tern of bootstrap samples $\{y_1\}_{b}$, $\{y_2\}_{b}$ and $\{y_3\}_{b}$, for $b = 1, \ldots, B$, of sizes $n_1$, $n_2$ and $n_3$, respectively;
\item from the $b$-th tern of bootstrap samples, compute the estimate $\widehat{\gamma}_{b}$, and then, $\ell_{b}(\widehat{\gamma})$;
\item get the estimates $\widehat{\mathrm{Med}}(\ell(\gamma_0))$ as the sample medians from the values $\ell_{b}(\widehat{\gamma})$, $b = 1,\ldots, B$.
\end{enumerate}
Again, only those bootstrap samples whose sample averages respect the fixed ordering of the population means ($\mu_1 < \mu_2 < \mu_3$) are processed. The confidence interval for $\gamma_0$ is therefore obtained as
\[
{\cal R}_{\gamma,\alpha} = \left\{\gamma: \widehat{w} \ell(\gamma) \le \chi_{1, (1 - \alpha)}^2 \right\},
\]
where $\alpha \in (0,1)$ and $\chi_{1, (1 - \alpha)}^2$ is the $(1 - \alpha)$-th quantile of a $\chi_1^2$ distribution. The interval ${\mathcal{R}}_{\gamma,\alpha}$ will denoted as ELQB.

The technique proposed to build confidence intervals for the VUS 
can be easily extended to the case where ties are present in the samples. 
Our approach here does not require the distribution functions $F_1$, $F_2$, and $F_3$ to be continuous. When ties are present in the data, it is sufficient to use in (\ref{ell:vus}) the appropriate estimate of $\gamma$, i.e.,
\begin{eqnarray}
\widehat{\gamma} &=& \frac{1}{n_1 n_2 n_3}\sum_{i = 1}^{n_1} \sum_{j = 1}^{n_2} \sum_{k = 1}^{n_3} \biggl\{\I\left(y_{1i} < y_{2j} < y_{3k}\right) + \frac{1}{2} \I\left(y_{1i} = y_{2j} < y_{3k}\right) \nonumber \\
&& + \: \frac{1}{2}\I\left(y_{1i} < y_{2j} = y_{3k}\right) + \frac{1}{6}\I\left(y_{1i} = y_{2j} = y_{3k}\right) \biggr\}.
\label{est:vus_ties}
\end{eqnarray}

\subsection{Confidence regions for the pair {\bf (TCF$_2$},{\bf TCF$_3$)}}
\label{sec:2:4}
Suppose now that the researcher fixes the value $\theta_1$ for $\TCF_1$. Let $\widehat{t}_1 = \widehat{F}_1^{-1}(\theta_1)$. By using the plugin method, again, we have an estimated version of EL statistic $\ell(\theta_1, \theta_2, \theta_3; t_1, t_2)$ as follows:
\begin{eqnarray}
\ell_{**}(\theta_2, \theta_3; t_2) &=& \ell(\theta_1, \theta_2, \theta_3; \widehat{t}_1, t_2) \nonumber \\
&=& 2n_1 \left\{\widehat{F}_1(\widehat{t}_1) \log \frac{\widehat{F}_1(\widehat{t}_1)}{\theta_1} + \left[1 - \widehat{F}_1(\widehat{t}_1)\right] \log\frac{1 - \widehat{F}_1(\widehat{t}_1)}{1 - \theta_1} \right\} \nonumber \\ 
&& + \: 2n_2 \left\{\widehat{P}(\widehat{t}_1, t_2) \log \frac{\widehat{P}(\widehat{t}_1, t_2)}{\theta_2} + \left[1 - \widehat{P}(\widehat{t}_1, t_2)\right] \log\frac{1 - \widehat{P}(\widehat{t}_1, t_2)}{1 - \theta_2} \right\} \nonumber \\
&& + \: 2n_3 \left\{\widehat{F}_3(t_2) \log \frac{\widehat{F}_3(t_2)}{1 - \theta_3} + \left[1 - \widehat{F}_3(t_2)\right]\log\frac{1 - \widehat{F}_3(t_2)}{\theta_3} \right\}, 
\label{emp:like:2_t1}
\end{eqnarray}
given $\theta_1$, if
\[
\begin{cases}
\widehat{t}_1 \text{ or } t_2 \in \left[y_{2(1)}, y_{2(n_2)}\right), \\
t_2 \in \left[y_{3(1)}, y_{3(n_3)}\right),
\end{cases}
\]
otherwise $\ell_{**}(\theta_2, \theta_3; t_2) = +\infty$. The following theorem shows 
how to obtain from $\ell_{**}(\theta_2, \theta_3; t_2)$ regions for the true pair $(\theta_{20}, \theta_{30})$ at a fixed value $t_{20}$ for the second threshold, given $\theta_1=\theta_{10}$.

\begin{theorem}
Assume that $F_1$ and $F_2$ have continuous density functions $f_1$ and $f_2$, such that $f_1(t_{10}) > 0$ and $f_2(t_{10}) > 0$. Let $t_{10}=F_1^{-1}(\theta_{10})$. If $\min\{n_1, n_2, n_3\} \to +\infty$ and the ratio $n_1/n_2$ has finite non-zero limit, then
\begin{eqnarray}
\ell_{**}(\theta_{20}, \theta_{30}; t_{20}) = \ell(\theta_{10}, \theta_{20}, \theta_{30}; \widehat{t}_{10}, t_{20})
\stackrel{d}{\longrightarrow} w U_1 + U_2,
\label{emp:like:2_3_t1}
\end{eqnarray}
where $w > 0$ is a suitable finite parameter, $U_1$ and $U_2$ are independent $\chi_1^2$ random variables.
\end{theorem}

The proof can be found in Section E of the Supplementary Materials. From the results in the proof, to estimate $w$, we propose to use $\widehat{w} = \dfrac{\widehat{\mathrm{Med}}(\ell_2(\theta_{20}; \widehat{t}_{10}, t_{20}))}{(7/9)^3}$, with 
$$
\ell_2(\theta_{20}; \widehat{t}_{10}, t_{20})=
2n_2 \left\{\widehat{P}(\widehat{t}_{10}, t_{20}) \log \frac{\widehat{P}(\widehat{t}_{10}, t_{20})}{\theta_{20}} + \left[1 - \widehat{P}(\widehat{t}_{10}, t_{20})\right] \log\frac{1 - \widehat{P}(\widehat{t}_{10}, t_{20})}{1 - \theta_{20}} \right\}, 
$$ 
and $\widehat{\mathrm{Med}}(\ell_2(\theta_{20}; \widehat{t}_{10}, t_{20}))$ obtained by the following bootstrap procedure:
\begin{enumerate}
\item from observed data $y_{11},\ldots, y_{1n_1}$ and $y_{21},\ldots, y_{2n_2}$ obtain the estimates $\widehat{t}_{10}= \widehat{F}_1^{-1}(\theta_{10})$ and $\widehat{\theta}_{20} = \widehat{F}_2(t_{20}) - \widehat{F}_2(\widehat{t}_{10})$;
\item get $B$ bootstrap samples $\{y_1\}_{b}$ and $\{y_2\}_{b}$ for $b = 1, \ldots, B$, of sizes $n_1$ and $n_2$, respectively;
\item from the $b$-th pair of the bootstrap sample, compute the estimate $\widehat{t}_{10b}$, and then, $\ell_{2b}(\widehat{\theta}_{20}; \widehat{t}_{10b}, t_{20})$;
\item get the estimates $\widehat{\mathrm{Med}}(\ell_2(\theta_{20}; \widehat{t}_{10}, t_{20}))$ as the sample medians from the values $\ell_{2b}(\widehat{\theta}_{20}; \widehat{t}_{10b}, t_{20})$, $b = 1,\ldots, B$.
\end{enumerate}

Notes to the bootstrap procedure of Section~\ref{sec:2:2} apply also here. Then, fixed $\theta_1 = \theta_{10}$, a confidence region for the pair $(\theta_{20}, \theta_{30})$,  with nominal coverage $1 - \alpha$, is obtained as
\[
{\cal R}^*_{23,\alpha} = \left\{(\theta_2, \theta_3): \ell_{**}(\theta_{2}, \theta_{3}; t_{20}) \le \widehat{c}_{\alpha} \right\},
\]
where $\alpha \in (0,1)$ and $\widehat{c}_{\alpha}$ is the sample quantile of order $(1 - \alpha)$, from 1,000 Monte Carlo values generated as $\widehat{w} U_1 + U_2$, where $U_1$ and $U_2$ are independent $\chi_1^2$ random variables. The interval ${\mathcal{R}}_{23,\alpha}^*$ will be denoted as ELQB. Of course, the approach proposed here can also be used to construct confidence regions for a different pair of TCFs, such as $(\TCF_1, \TCF_3)$, fixed $\theta_2 = \theta_{20}$. 

\section{Simulation study}
\label{sec:simu}

\subsection{Simulation set-up}
\label{sec:simu:1}
To investigate the finite sample behaviour of our proposed empirical likelihood techniques, based on $\ell(\theta_1, \theta_2, \theta_3; t_1, t_2)$ in (\ref{emp:like:3C}), $\ell_*(\theta_2)$ in (\ref{emp:like:2_2}), $l(\gamma)$ in (\ref{ell:vus}) and $l_{**}(\theta_2,\theta_3;t_2)$ in (\ref{emp:like:2_t1}), we conducted a large simulation study. In particular, we evaluated the coverage probability of 3D confidence regions for the triplet of TCFs $(\theta_{10}, \theta_{20}, \theta_{30})$, at fixed thresholds $(t_{10}, t_{20})$, of confidence intervals for $\TCF_2$, $\theta_{20}$, at fixed $\TCF_1$ and $\TCF_3$, of confidence intervals for the VUS and confidence regions for the pair $(\TCF_2, \TCF_3)$, $(\theta_{20},\theta_{30})$, at fixed $\TCF_1$. We also compared our proposals with competitors, when present in the literature.

In the simulation experiments, we considered ten scenarios, listed in Table \ref{tab:1}. In particular, the first three scenarios refer to the tri-normal setting; scenarios 4 to 6 refer to the case of mixed distributions (gamma, log-normal, and Weibull distributions); scenarios 7 to 9 consider a tri-beta setting, where the biomarkers' values are bounded in (0, 1); the last scenario refers to a setting that considers mixture distributions.

\begin{center}
[[Table 1. about here]]
\end{center}

\subsection{Results: confidence regions for TCFs}
\label{sec:simu:2}
The performance of our proposed empirical likelihood confidence region ELQ3D is evaluated at three different levels of nominal coverage $1 - \alpha$, i.e., 0.90, 0.95 and 0.99. Under each scenario in Table \ref{tab:1}, 10,000 random samples are generated. The sample sizes $(n_1, n_2, n_3)$ are set as (30, 30, 30), (50, 50, 50) and (100, 100, 100). The simulation results are presented in Table \ref{tab:2}.

\begin{center}
[[Table 2. about here]]
\end{center}

As one can see, the empirical coverages are close to the nominal ones in almost all the considered settings. As expected, our empirical likelihood confidence region ELQ3D needs a larger sample size when the true TCFs are close to 1.

\subsection{Results: confidence intervals for {\rm TCF}$_2$, at fixed $\theta_{10}$ and $\theta_{30}$}
\label{sec:simu:3}
Here, we compare the performance of our proposed methods ELQB in Section (\ref{sec:2:2}) with the existing nonparametric approaches, i.e., ELP, ELB, and BTII by \citet{dong2015confidence}, IF by \citet{hai2022bayesian}, PEL and AEL by \citet{rahman2022empirical}, through scenarios in Table \ref{tab:1}. Under each scenario, we fixed the values $\theta_{10}$ and $\theta_{30}$, 
and generated 5,000 random samples. The sample sizes $(n_1, n_2, n_3)$ are set at (30, 30, 30), (50, 30, 30), (50, 50, 50), (100, 50, 50), (100, 100, 50) and (100, 100, 100). For, the ELB and BTII methods, we consider 500 bootstrap samples, whereas, for our methods, we use $B=200$. Simulation results are reported in Tables \ref{tab:3} - \ref{tab:6}. Although BTII is not an EL-based method, it is considered here as a reference, due to its nonparametric nature and its relative ease of use.

Overall, results in Tables \ref{tab:3} - \ref{tab:6} (and Table S1, Section F of the Supplementary Materials) seem to show that the proposed method, ELQB, competes with the best contestants (ELB, BTII),  arriving at outperforming in some scenarios, in particular in scenario 10, with mixture models (see Table \ref{tab:6}). In general, the approaches IF, AEL, and PEL behave poorly. The IF approach requires greater sample sizes (than that of other approaches) in some scenarios (3, Table \ref{tab:3} and 6, Table \ref{tab:4}). The AEL and PEL approaches seem, sometimes, not to yield consistent results even when $n$ grows (see scenario 3 in Table S1, Section F of the Supplementary Materials). Moreover, the three methods 
seem not to be able to cope with mixture models (Table \ref{tab:6}). 

Of course, fully nonparametric approaches, such as those here considered, typically perform well only with large sample sizes when true TCFs values are close to 1. In such situations,  a sample size of at least greater than 100 for each class is required.

\begin{center}
[[Table 3. about here]]
\end{center}

\begin{center}
[[Table 4. about here]]
\end{center}

\begin{center}
[[Table 5. about here]]
\end{center}

\begin{center}
[[Table 6. about here]]
\end{center}

\subsection{Results: confidence intervals for the VUS}
\label{sec:simu:4}
We examine the performance of our proposed EL-based method ELQB in Section (\ref{sec:2:3}), for constructing confidence intervals for the VUS. We also compare our method to the existing approaches, ELU \citep{wan2012empirical} and JEL \citep{guangming2013nonparametric}, through scenarios in Table \ref{tab:1}. Under each scenario, we generated 5,000 random samples. The sample sizes $(n_1, n_2, n_3)$ are set at (15, 15, 15), (30, 30, 30), (50, 30, 30), (50, 50, 50), (75, 75, 75) and (100, 100, 100). Simulation results are reported in Tables \ref{tab:vus:1} - \ref{tab:vus:4}. We also considered some additional scenarios, where the true VUS value is around 0.45 to 0.55; corresponding simulation results are given in Table S2, Section G, of the Supplementary Materials. 

Overall, our approach seems to perform well and is often more accurate than competitors, in all scenarios, particularly when the VUS's true value is large. The ELU method seems to be the least accurate.

\subsection{Results: confidence regions for the pair $(\TCF_2,\TCF_3)$, at fixed $\theta_{10}$}
\label{sec:simu:5}
Finally, Table 11 and Tables S3-S5 in Section H of the Supplementary Materials report simulation results about confidence regions for the pairs $(\TCF_2,\TCF_3)$, at fixed $\theta_1=\theta_{10}$, build using our approach described in Section (\ref{sec:2:4}). In such simulation experiments, we consider 5,000  Monte Carlo replication and some values for the true class fractions in each scenario. Again we set $B = 200$.

As one can see, our method performs well in all considered cases, with the need (as expected) for larger sample sizes (at least (50, 50, 50)) as the true values of the TCFs approach 1.

\begin{center}
[[Table 7. about here]]
\end{center}

\begin{center}
[[Table 8. about here]]
\end{center}

\begin{center}
[[Table 9. about here]]
\end{center}

\begin{center}
[[Table 10. about here]]
\end{center}

\begin{center}
[[Table 11, about here]]
\end{center}

\section{An illustrative example}
\label{sec:apply}
In this section, we use a genomic dataset and apply our proposed methods to evaluate the ability of some gene expressions to distinguish ductal carcinomas in situ (DCIS) from noncancerous (NC) and invasive breast cancers (IBC). We consider the raw data from series record GSE214540, published on the GEO repository by \citet{guvakova2022g3mclass}. The raw data contains mRNA expression levels of different genes from FFPE (formalin-fixed paraffin-embedded) human breast tissue samples generated by using the QuantiGene Plex 2.0 assay and Flex-Map 3D \citep{prabakaran2019gaussian}. The FFPE tissues were collected from the Department of Pathology and Laboratory Medicine, Tumor Tissue and Biospecimen Bank, and the Cooperative Human Tissue Network at the University of Pennsylvania. 

\citet{guvakova2022g3mclass} measured 14 target genes, namely: ESR1 (estrogen receptor 1), PGR (progesterone receptor), ERBB2 (erb-b2 receptor tyrosine kinase 2), IGF1R (insulin-like growth factor 1 receptor), VAV1 (vav guanine nucleotide exchange factor 1), VAV2 (vav guanine nucleotide exchange factor 2), VAV3 (vav guanine nucleotide exchange factor 3), RAP1A (ras-related protein Rap-1A), RAP1B (ras-related protein Rap-1b), RAPGEF1 (Rap guanine nucleotide exchange factor 1), KRT5 (keratin 5), KRT8 (keratin 8), CDH1 (cadherin 1), CDH2 (cadherin 2); and two housekeeping genes, namely, PPIB (peptidylprolyl isomerase B) and GUSB (glucuronidase beta). As noted by \citet{prabakaran2019gaussian}, in FFPE tissue, PPIB expressed consistently, whereas GUSB showed a relatively low expression level. For this reason, we do not take into account the mRNA expression levels of GUSB in our analysis.

Before doing the analysis, the normalization of the raw mRNA data is required. Within our analysis, the normalized data are obtained through three steps: firstly, we subtract background values from each measurement, to obtain the real values of gene expressions; secondly, we obtain the housekeeping normalization factor by dividing the average of housekeeping gene values, the PPIB, to each of its values; then, the normalized mRNA values of each gene is obtained as a ratio of mRNA values and the housekeeping normalization factor. Note that, in the first step,  negative values are set as 0, to reflect the fact that there is no measurable expression of the target gene in the assay. We then apply $\log_{10}(x + 1)$ transformation to the normalized values for the main analysis.

The final dataset contains the $\log_{10}$-transformed (normalized) mRNA expression values of 14 target genes from FFPE tissues of 251 women: 34 for the NC group, 75 for the DCIS group, and 142 for the IBC group. A preliminary analysis indicates that almost all genes have poor accuracy in classifying the three stages of breast cancer (see VUS estimates in Table S6, Section I of the Supplementary Materials). Hence, we consider a linear combinations, say $T$, of the $\log_{10}$-transformed mRNA expression levels for three genes: KRT5, KRT8 and CDH2, i.e., $T = \text{KRT8} - 0.9 \times \text{KRT5} + 0.3 \times \text{CDH2}$. The coefficients of such a combination are based on the maximization process of the estimated VUS, as in \citet{Zhang2011}. Then, in the analysis, we treat this combination as exogenously fixed.

The estimated VUS for the combined test $T$ is 0.685. Employing our method ELBQ, proposed in Section \ref{sec:2:3}, the 90\%, 95\%  and 99\% EL confidence intervals for the VUS are $(0.626, 0.740)$, $(0.614, 0.750)$ and $(0.591, 0.769)$, respectively. By these results, $T$ seems to have a sufficiently good capacity for discrimination among the three stages of breast cancer.

\begin{center}
[[Figure 1. about here]]
\end{center}

\begin{figure}[!p]
\begin{center}
\includegraphics[width=0.5\textwidth]{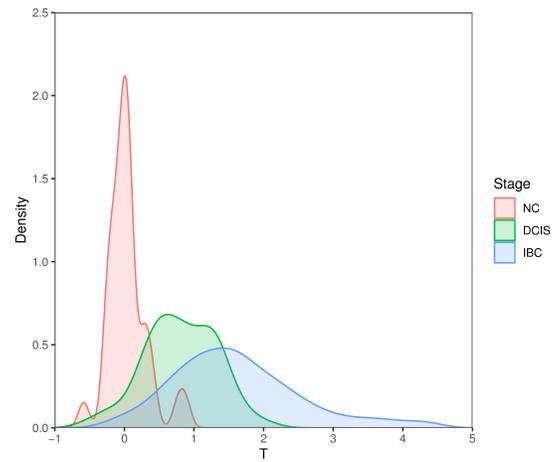}
\end{center}
\caption{Estimated densities for the combination $T = \text{KRT8} - 0.9 \times \text{KRT5} + 0.3 \times \text{CDH2}$.}
\label{fig:densities:T}
\end{figure}

\begin{figure}[!p]
\begin{center}
\includegraphics[width=0.5\textwidth, clip, trim = 3cm 1.5cm 1cm 5cm]{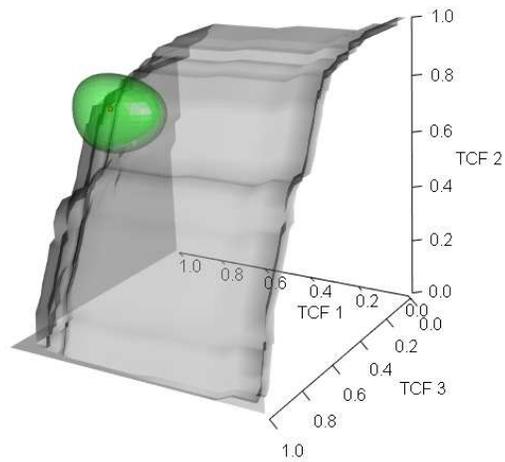}
\end{center}
\caption{Empirical ROC surface for $T$, and the confidence region for $(\theta_{10}, \theta_{20}, \theta_{30})$ when the pair $(t_1, t_2)$ is $(0.275, 1.35)$. The point estimate $(\widehat{\theta}_{1}, \widehat{\theta}_{2}, \widehat{\theta}_{3})$ is $(0.824, 0.747, 0.578)$.}
\label{fig:roc_surface}
\end{figure}

\begin{figure}[!p]
\begin{center}
\includegraphics[width=0.5\textwidth]{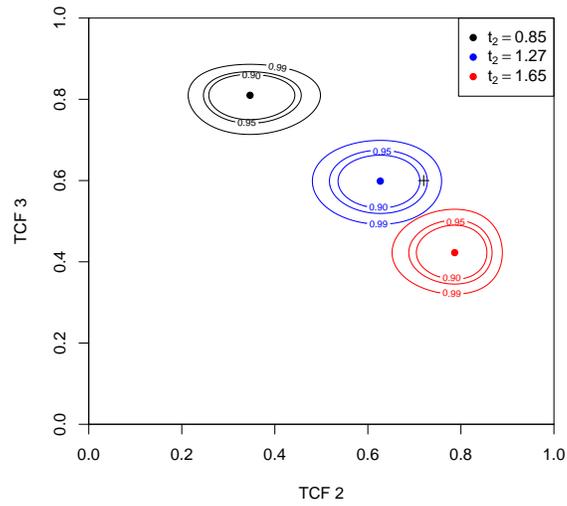}
\end{center}
\caption{Confidence regions for $(\theta_{20}, \theta_{30})$ when $\theta_{10} = 0.9$, for three different choices of $t_{2}$: 0.85, 1.27 and 1.65. The point estimates $(\widehat{\theta}_{2}, \widehat{\theta}_{3})$ are $(0.347, 0.810)$, $(0.627, 0.599)$ and $(0.787, 0.423)$, respectively. Symbol ``+'' is at $\TCF_2 = 0.72$ and $\TCF_3 = 0.6$.}
\label{fig:cr:tcf23}
\end{figure}

For the three stages, the kernel-based estimated densities of $T$ are shown in Figure \ref{fig:densities:T}. By inspection of the figure, we choose two plausible values for the thresholds $t_1$ and $t_2$: 0.275 and 1.35, respectively. Treating these values as fixed, Figure \ref{fig:roc_surface} shows the corresponding 95\% confidence region for the TCFs $(\theta_{10}, \theta_{20}, \theta_{30})$ 
on the estimated ROC surface, obtained by our ELQ3D method in Section \ref{sec:2:1}. Moreover, if we fix $\theta_{10} = 0.8$ and $\theta_{30} = 0.6$, the empirical estimate $\widehat{\theta}_2$ is about 0.707, and the 95\% ELQB (Section \ref{sec:2:2}, with 200 bootstrap replications) confidence interval for the probability of correct classification in the DCIS stage
is $(0.461, 0.889)$. The width of such an interval reflects the variability of the data and the degree of overlap among the estimated densities of the combination $T$, in particular between the DCIS and IBC stages. Finally, Figure \ref{fig:cr:tcf23} shows ELQB confidence regions (Section \ref{sec:2:4}, with 200 bootstrap replications) for the pair $(\theta_{20}, \theta_{30})$ when $\theta_{10} = 0.9$ 
and the threshold $t_2$ is chosen to be 0.85, 1.27, and 1.65, respectively. Confidence regions indicate, at different levels, the pairs $(\theta_{2}, \theta_{3})$ which are compatible with the constraint $\theta_{10} = 0.9$ for the combination $T$, for three possible choices of threshold $t_2$. Then, for example, Figure \ref{fig:cr:tcf23} indicates that, at level 0.95, the combination $T$ may perform with values for $\TCF_2$ and $\TCF_3$ equal to 0.72 and 0.6, respectively when we require $\TCF_1 = 0.9$ and use $t_2 = 1.27$.

\begin{center}
[[Figure 2. about here]]
\end{center}

\begin{center}
[[Figure 3. about here]]
\end{center}

\section{Discussion}
\label{sec:disc}
We present a fairly general approach for constructing confidence intervals and regions in a three-class ROC analysis. Our approach allows getting adequate techniques to solve inferential problems concerning the evaluation of a diagnostic test (or a biomarker) and, in particular, to obtain: (i) confidence regions for the triplet $(\TCF_1, \TCF_2, \TCF_3)$ corresponding to a specific choice for the thresholds $t_1$ and $t_2$; (ii) confidence intervals for the VUS; (iii) confidence intervals for the probability of correct classification to the ``early stage'', $\TCF_2$, for fixed $\TCF_1$ and $\TCF_3$; (iv) confidence regions for a pair of TCFs, when it is fixed the value of the remaining third. The proposed techniques are justified theoretically and are validated empirically using a large simulation study, where they are also compared with competitors, if present in the literature. Overall, simulation results reveal that the proposed methods perform well in general, and are at least as accurate as competitors, showing better performance in several situations. 

We use a genomics dataset to illustrate an application of our techniques. However, the proposed approach can be also easily extended to provide solutions for inferential problems other than those treated and mentioned above. Without going into details, we discuss some cases here.

\textbf{Confidence intervals for the HUM.} Suppose that the disease is articulated according to $M > 3$ ordered stages. Let $Y_j$, $j = 1, \ldots, M$, be the test results for a subject within $j$-th class. In such a situation, the hypervolume under the ROC manifold (HUM), extends the concept of VUS and is defined as $\beta = \Pr(Y_1 < Y_2 < \ldots < Y_M)$. An estimator, consistent and asymptotically normal \citep{nak:04}, of $\beta$ can be obtained as:
\[
\widehat{\beta} = \widehat{\Pr}(Y_1 < Y_2 < \ldots < Y_M) = \frac{1}{n_1 n_2 \ldots n_M}\sum_{i = 1}^{n_1} \sum_{r = 1}^{n_2} \ldots \sum_{k = 1}^{n_M} \I\left({Y}_{1i} < {Y}_{2r} < \ldots < {Y}_{Mk} \right).
\]
Since $\beta$ is a probability, we can write the empirical likelihood statistic $\ell(\beta)$ as in (\ref{ell:vus}), and by generalizing Theorem \ref{thrm:vus}, can prove that $\ell(\beta_0)$ has asymptotically a scaled $\chi^2_1$ distribution, under some weak conditions. Thus, a confidence interval for the HUM can be obtained by
\[
{\cal R}_{\beta,\alpha} = \left\{\beta: \widehat{w} \ell(\beta) \le \chi_{1, (1 - \alpha)}^2 \right\}.
\]

\textbf{Confidence intervals for the covariate-specific VUS.} Often, in biomedical studies, the researchers collect not only results of potential diagnostic tests but also additional information, about the subjects under study, as covariates (e.g., age, sex, comorbidity profiles). In such cases, covariate-specific measures to evaluate the accuracy of the diagnostic tests are relevant. For the three-class setting, covariate-specific VUS estimators are proposed in \citet{to2022estimation}. Such estimators are consistent and asymptotically normal. If, for a given vector of covariate values $x$, $\widehat{\gamma}(x)$ denotes an estimate, a confidence interval for the covariate-specific VUS, $\gamma(x)$, can be obtained again by (\ref{ell:vus}), where $\gamma$ and $\widehat{\gamma}$ are replaced by $\gamma(x)$ and $\widehat{\gamma}(x)$, respectively.

\textbf{Confidence regions for optimal thresholds and associated TCF.} Although scarcely used, a criterion for choosing an optimal threshold in a two-class setting, is the so-called symmetric point (see \citealp{lopez2016confidence} and references therein). This approach finds the threshold $t$ at which the sensitivity and specificity of the diagnostic test have the same value. In the three-class setting, the extension of the symmetric point approach is trivial: the optimal thresholds $t_1$, $t_2$ are such that $\TCF_1$, $\TCF_2$ and $\TCF_3$ have the same value, i.e., $\theta_1 = \theta_2 = \theta_3$. Because this criterion imposes two constraints on $\theta_1$, $\theta_2$ and $\theta_3$, i.e., $\theta_1 = \theta_2$ and $\theta_2 = \theta_3$, if we denote by $\theta$ the common value and, from (\ref{emp:like:3C}), let  $\ell_+(\theta, t_1, t_2) = \ell(\theta, \theta, \theta; t_1, t_2)$, we can prove that $\ell_+(\theta, t_1, t_2)$ again has an asymptotic $\chi^2_3$ distribution, under the true parameters values. Therefore, $\ell_+$ can be used to build confidence regions for the symmetric point-based optimal thresholds and the associated common value of TCFs. This technique extends the proposal discussed in \citet{adimari2020nonparametric} for two classes, to the three-class setting. 

An interesting topic that remains to be developed concerns the problem of building confidence regions for optimal TCFs, i.e., TCFs corresponding to thresholds chosen through other criteria, such as the one based on the generalized Youden index \citep{nakas2010accuracy}, the closest to perfection and the max volume \citep{attwood2014diagnostic}. Such a topic will be the focus of future work.

\section{Software}
\label{sec:software}
Software in the form of R codes is available on \url{https://github.com/toduckhanh/emplikROCS}

\section*{Supplementary Material}
\label{sec6}

Supplementary material is available online at \url{http://biostatistics.oxfordjournals.org}. 

\section*{Acknowledgments}

\bibliographystyle{biorefs}
\bibliography{refs_empi_3classes_Bios}

\begin{thebibliography}{99}

\bibitem[Adimari(1998)Adimari]{adimari1998empirical}
\textsc{Adimari, G.} (1998).
\newblock An empirical likelihood statistic for quantiles.
\newblock {\em Journal of Statistical Computation and
  Simulation\/}~\textbf{60}(1), 85--95.

\bibitem[Adimari and Guolo(2010)Adimari and Guolo]{adimari2010note}
\textsc{Adimari, G. and Guolo, A.} (2010).
\newblock A note on the asymptotic behaviour of empirical likelihood
  statistics.
\newblock {\em Statistical Methods \& Applications\/}~\textbf{19}, 463--476.

\bibitem[Adimari and Sinigaglia(2020)Adimari and
  Sinigaglia]{adimari2020nonparametric}
\textsc{Adimari, G. and Sinigaglia, A.} (2020).
\newblock Nonparametric confidence regions for the symmetry point-based optimal
  cutpoint and associated sensitivity of a continuous-scale diagnostic test.
\newblock {\em Biometrical Journal\/}~\textbf{62}(6), 1463--1475.

\bibitem[Attwood \emph{and others}(2014)Attwood, Tian and
  Xiong]{attwood2014diagnostic}
\textsc{Attwood, K., Tian, L. and Xiong, C.} (2014).
\newblock Diagnostic thresholds with three ordinal groups.
\newblock {\em Journal of Biopharmaceutical Statistics\/}~\textbf{24}(3),
  608--633.

\bibitem[Dai and Luke(2008)Dai and Luke]{misc3d}
\textsc{Dai, F. and Luke, T.} (2008).
\newblock Computing and displaying isosurfaces in {R}.
\newblock {\em Journal of Statistical Software\/}~\textbf{28}(1), 1--14.

\bibitem[Dong and Tian(2015)Dong and Tian]{dong2015confidence}
\textsc{Dong, T. and Tian, L.} (2015).
\newblock Confidence interval estimation for sensitivity to the early diseased
  stage based on empirical likelihood.
\newblock {\em Journal of Biopharmaceutical Statistics\/}~\textbf{25}(6),
  1215--1233.

\bibitem[Dong \emph{and others}(2011)Dong, Tian, Hutson and
  Xiong]{dong2011parametric}
\textsc{Dong, T., Tian, L., Hutson, A. and Xiong, C.} (2011).
\newblock Parametric and non-parametric confidence intervals of the probability
  of identifying early disease stage given sensitivity to full disease and
  specificity with three ordinal diagnostic groups.
\newblock {\em Statistics in Medicine\/}~\textbf{30}(30), 3532--3545.

\bibitem[Guangming \emph{and others}(2013)Guangming, Xiping and
  Wang]{guangming2013nonparametric}
\textsc{Guangming, P., Xiping, W. and Wang, Z.} (2013).
\newblock Nonparametric statistical inference for {$P(X < Y < Z)$}.
\newblock {\em Sankhya A\/}~\textbf{75}, 118--138.

\bibitem[Guilbaud(1980)Guilbaud]{guilbaud1980asymptotic}
\textsc{Guilbaud, O.} (1980).
\newblock Asymptotic behavior of the empirical distribution function at a
  random point and some applications.
\newblock {\em Scandinavian Journal of Statistics\/}~\textbf{7}(4), 181--189.

\bibitem[Guvakova and Sokol(2022)Guvakova and Sokol]{guvakova2022g3mclass}
\textsc{Guvakova, M.~A. and Sokol, S.} (2022).
\newblock The g3mclass is a practical software for multiclass classification on
  biomarkers.
\newblock {\em Scientific Reports\/}~\textbf{12}(1), 18742.

\bibitem[Hai \emph{and others}(2023)Hai, Shi and Qin]{hai2022bayesian}
\textsc{Hai, Y., Shi, S. and Qin, G.} (2023).
\newblock Bayesian and influence function-based empirical likelihoods for
  inference of sensitivity to the early diseased stage in diagnostic tests.
\newblock {\em Biometrical Journal\/}~\textbf{65}(3), 2200021.

\bibitem[Hjort \emph{and others}(2009)Hjort, McKeague and van
  Keilegom]{hjort2009extending}
\textsc{Hjort, N.~L., McKeague, I.~W. and van Keilegom, I.} (2009).
\newblock Extending the scope of empirical likelihood.
\newblock {\em Annals of Statistics\/}~\textbf{37}(3), 1079--1111.

\bibitem[Jing \emph{and others}(2009)Jing, Yuan and Zhou]{jing2009jackknife}
\textsc{Jing, B.~Y., Yuan, J. and Zhou, W.} (2009).
\newblock Jackknife empirical likelihood.
\newblock {\em Journal of the American Statistical
  Association\/}~\textbf{104}(487), 1224--1232.

\bibitem[Kang and Tian(2013)Kang and Tian]{kang2013estimation}
\textsc{Kang, L. and Tian, L.} (2013).
\newblock Estimation of the volume under the roc surface with three ordinal
  diagnostic categories.
\newblock {\em Computational Statistics \& Data Analysis\/}~\textbf{62},
  39--51.

\bibitem[Lazar(2021)Lazar]{lazar2021review}
\textsc{Lazar, N.~A.} (2021).
\newblock A review of empirical likelihood.
\newblock {\em Annual Review of Statistics and its Application\/}~\textbf{8},
  329--344.

\bibitem[Lehmann(1998)Lehmann]{lehmann1998elements}
\textsc{Lehmann, E.~L.} (1998).
\newblock {\em Elements of Large-Sample Theory\/}. Springer Science \& Business
  Media.

\bibitem[Li and Zhou(2009)Li and Zhou]{li2009nonparametric}
\textsc{Li, J. and Zhou, X.~H.} (2009).
\newblock Nonparametric and semiparametric estimation of the three way receiver
  operating characteristic surface.
\newblock {\em Journal of Statistical Planning and
  Inference\/}~\textbf{139}(12), 4133--4142.

\bibitem[Liu and Zhao(2022)Liu and Zhao]{liu2022review}
\textsc{Liu, P. and Zhao, Y.} (2022).
\newblock A review of recent advances in empirical likelihood.
\newblock {\em Wiley Interdisciplinary Reviews: Computational Statistics\/},
  e1599.

\bibitem[L{\'o}pez-Rat{\'o}n \emph{and others}(2016)L{\'o}pez-Rat{\'o}n,
  Cadarso-Su{\'a}rez, Molanes-L{\'o}pez and Let{\'o}n]{lopez2016confidence}
\textsc{L{\'o}pez-Rat{\'o}n, M., Cadarso-Su{\'a}rez, C., Molanes-L{\'o}pez,
  E.~M. and Let{\'o}n, E.} (2016).
\newblock Confidence intervals for the symmetry point: an optimal cutpoint in
  continuous diagnostic tests.
\newblock {\em Pharmaceutical Statistics\/}~\textbf{15}(2), 178--192.

\bibitem[Nakas(2014)Nakas]{nak:14}
\textsc{Nakas, C.~T.} (2014).
\newblock Developments in {ROC} surface analysis and assessment of diagnostic
  markers in three-class classification problems.
\newblock {\em REVSTAT{-}Statistical Journal\/}~\textbf{12}(1), 43--65.

\bibitem[Nakas \emph{and others}(2010)Nakas, Alonzo and
  Yiannoutsos]{nakas2010accuracy}
\textsc{Nakas, C.~T., Alonzo, T.~A. and Yiannoutsos, C.~T.} (2010).
\newblock Accuracy and cut-off point selection in three-class classification
  problems using a generalization of the {Y}ouden index.
\newblock {\em Statistics in Medicine\/}~\textbf{29}(28), 2946--2955.

\bibitem[Nakas and Yiannoutsos(2004)Nakas and Yiannoutsos]{nak:04}
\textsc{Nakas, C.~T. and Yiannoutsos, C.~T.} (2004).
\newblock Ordered multiple-class {ROC} analysis with continuous measurements.
\newblock {\em Statistics in Medicine\/}~\textbf{23}(22), 3437--3449.

\bibitem[Owen(2001)Owen]{owen2001empirical}
\textsc{Owen, A.~B.} (2001).
\newblock {\em Empirical likelihood\/}. Chapman and Hall/CRC.

\bibitem[Prabakaran \emph{and others}(2019)Prabakaran, Wu, Lee, Tong, Steeman,
  Koo, Zhang and Guvakova]{prabakaran2019gaussian}
\textsc{Prabakaran, I., Wu, Z., Lee, C., Tong, B., Steeman, S., Koo, G., Zhang,
  P.~J. and Guvakova, M.~A.} (2019).
\newblock Gaussian mixture models for probabilistic classification of breast
  cancergaussian mixture models for cancer classification.
\newblock {\em Cancer Research\/}~\textbf{79}(13), 3492--3502.

\bibitem[{R Core Team}(2022){R Core Team}]{R_team}
\textsc{{R Core Team}}. (2022).
\newblock {\em R: A Language and Environment for Statistical Computing\/}.
\newblock R Foundation for Statistical Computing, Vienna, Austria.
\newblock URL https://www.R-project.org/.

\bibitem[Rahman \emph{and others}(2022)Rahman, Zhao and
  Initiative]{rahman2022empirical}
\textsc{Rahman, H., Zhao, Y. and Initiative, Alzheimer's Disease~Neuroimaging}.
  (2022).
\newblock Empirical likelihood confidence interval for sensitivity to the early
  disease stage.
\newblock {\em Pharmaceutical Statistics\/}~\textbf{21}(3), 566--583.

\bibitem[To \emph{and others}(2022)To, Adimari and Chiogna]{to2022estimation}
\textsc{To, D.-K., Adimari, G. and Chiogna, M.} (2022).
\newblock Estimation of the volume under a {ROC} surface in presence of
  covariates.
\newblock {\em Computational Statistics \& Data Analysis\/}~\textbf{174},
  107434--107448.

\bibitem[Wan(2012)Wan]{wan2012empirical}
\textsc{Wan, S.} (2012).
\newblock An empirical likelihood confidence interval for the volume under
  {ROC} surface.
\newblock {\em Statistics \& Probability Letters\/}~\textbf{82}(7), 1463--1467.

\bibitem[Xiong \emph{and others}(2006)Xiong, van Belle, Miller and
  Morris]{xiong2006measuring}
\textsc{Xiong, C., van Belle, G., Miller, J.~P. and Morris, J.~C.} (2006).
\newblock Measuring and estimating diagnostic accuracy when there are three
  ordinal diagnostic groups.
\newblock {\em Statistics in Medicine\/}~\textbf{25}(7), 1251--1273.

\bibitem[Zhang and Li(2011)Zhang and Li]{Zhang2011}
\textsc{Zhang, Y. and Li, J.} (2011).
\newblock Combining multiple markers for multi-category classification: an
  {ROC} surface approach.
\newblock {\em Australian and New Zealand Journal of
  Statistics\/}~\textbf{53}(1), 63--78.

\end{thebibliography}

\begin{table}[!p]
\caption{Scenarios for the simulation study. Here, $\mathcal{N}$,
$\mathcal{LN}$, $\mathcal{G}$ and $\mathcal{W}$ indicate normal, log-normal, gamma and Weibull distributions; $\theta_{10}$, $\theta_{20}$ and $\theta_{30}$ are true values of TCFs; and $\gamma_0$ is the true value of VUS.}
\label{tab:1}
\begin{small}
\begin{center}
\begin{tabular}{@{}l r r r r r r r r r@{}}
\tblhead{
Scenario & $Y_1$ & $Y_2$ & $Y_3$ & $t_{10}$ & $t_{20}$ & $\theta_{10}$ & $\theta_{20}$ & $\theta_{30}$ & $\gamma_0$}
1 & $\mathcal{N}(0, 1)$ & $\mathcal{N}(2.5, 1.1^2)$ & $\mathcal{N}(3.69, 1.2^2)$ & 0.842 & 2.680 & 0.8 & 0.5 & 0.8 & 0.772 \\
2 & $\mathcal{N}(0, 1)$ & $\mathcal{N}(3.5, 1.1^2)$ & $\mathcal{N}(5.5, 1.2^2)$ & 0.842 & 4.490 & 0.8 & 0.8 & 0.8 & 0.881 \\
3 & $\mathcal{N}(0, 1)$ & $\mathcal{N}(4, 1.2^2)$ & $\mathcal{N}(8.189, 2^2)$ & 1.282 & 5.626 & 0.9 & 0.9 & 0.9 & 0.959 \\
4 & $\mathcal{G}(6, 12)$ & $\mathcal{LN}(1.5, 0.5)$ & $\mathcal{W}(4, 6.6)$ & 0.659 & 4.536 & 0.8 & 0.5 & 0.8 & 0.669\\
5 & $\mathcal{G}(6, 12)$ & $\mathcal{LN}(1.5, 0.5)$ & $\mathcal{W}(4, 10)$ & 0.659 & 6.873 & 0.8 & 0.8 & 0.8 & 0.868 \\
6 & $\mathcal{G}(6, 12)$ & $\mathcal{LN}(1.5, 0.5)$ & $\mathcal{W}(4, 12.4)$ & 0.659 & 8.523 & 0.8 & 0.9 & 0.8 & 0.927 \\
7 & $\mathcal{B}(1, 6)$ & $\mathcal{B}(6, 6)$ & $\mathcal{B}(9.6, 6)$ & 0.235 & 0.513 & 0.8 & 0.5 & 0.8 & 0.698 \\
8 & $\mathcal{B}(1, 6)$ & $\mathcal{B}(9, 6)$ & $\mathcal{B}(20.4, 6)$ & 0.235 & 0.707 & 0.8 & 0.8 & 0.8 & 0.869 \\
9 & $\mathcal{B}(1, 6)$ & $\mathcal{B}(6, 6)$ & $\mathcal{B}(20.4, 6)$ & 0.235 & 0.707 & 0.8 & 0.9 & 0.8 & 0.917 \\
10 & $0.5N(-1, 1) + $ & $0.5N(1, 1) + $ & $0.5N(3, 1.5) +$ & 0.5 & 4.5 & 0.5 & 0.674 & 0.522 & 0.544 \\
& $0.5N(2, 1)$ & $0.5N(4, 1.5)$ & $0.5N(6, 1)$ &  &  &  &  &  & \\
\lastline
\end{tabular}
\end{center}
\end{small}
\end{table}

\begin{table}[!p]
\caption{Monte Carlo coverages for the ELQ3D confidence regions for $(\theta_{10},\theta_{20}, \theta_{30})$, at fixed values for $t_1$ and $t_2$, for each scenario in Table 1.}
\label{tab:2}
\begin{footnotesize}
\begin{center}
\begin{tabular}{@{}r r r r c r r r@{}}
\tblhead{
Scenario & $\theta_{10}$ & $\theta_{20}$ & $\theta_{30}$ & $n_1 = n_2 = n_3$ & \multicolumn{3}{c}{Nominal level} \\
& & & & & 0.90 & 0.95 & 0.99}
\multirow{3}{1cm}{1} & & & & 30 & 0.892 & 0.944 & 0.988 \\
& 0.8 & 0.5 & 0.8 & 50 & 0.902 & 0.950 & 0.991 \\
& & & & 100 & 0.896 & 0.943 & 0.990
\lastline
\multirow{3}{1cm}{2} & & & & 30 & 0.890 & 0.939 & 0.988 \\
& 0.8 & 0.8 & 0.8 & 50 & 0.896 & 0.945 & 0.980 \\
& & & & 100 & 0.898 & 0.947 & 0.989 
\lastline
\multirow{3}{1cm}{3} & & & & 30 & 0.833 & 0.859 & 0.874 \\
& 0.9 & 0.9 & 0.9 & 50 & 0.895 & 0.945 & 0.986 \\
& & & & 100 & 0.895 & 0.949 & 0.990 
\lastline
\multirow{3}{1cm}{4} & & & & 30 & 0.903 & 0.951 & 0.989 \\
& 0.8 & 0.5 & 0.8 & 50 & 0.904 & 0.950 & 0.991 \\
& & & & 100 & 0.902 & 0.950 & 0.990 
\lastline
\multirow{3}{1cm}{5} & & & & 30 & 0.894 & 0.943 & 0.989 \\
& 0.8 & 0.8 & 0.8 & 50 & 0.897 & 0.947 & 0.990 \\
& & & & 100 & 0.896 & 0.944 & 0.987 
\lastline
\multirow{3}{1cm}{6} & & & & 30 & 0.866 & 0.910 & 0.946 \\
& 0.8 & 0.9 & 0.8 & 50 & 0.895 & 0.945 & 0.986 \\
& & & & 100 & 0.895 & 0.949 & 0.990 
\lastline
\multirow{3}{1cm}{7} & & & & 30 & 0.893 & 0.947 & 0.988 \\
& 0.8 & 0.5 & 0.8 & 50 & 0.897 & 0.946 & 0.990 \\
& & & & 100 & 0.897 & 0.947 & 0.990 
\lastline
\multirow{3}{1cm}{8} & & & & 30 & 0.894 & 0.943 & 0.988 \\
& 0.8 & 0.8 & 0.8 & 50 & 0.894 & 0.947 & 0.988 \\
& & & & 100 & 0.898 & 0.947 & 0.990
\lastline
\multirow{3}{1cm}{9} & & & & 30 & 0.867 & 0.911 & 0.948 \\
& 0.8 & 0.9 & 0.8 & 50 & 0.897 & 0.947 & 0.984 \\
& & & & 100 & 0.899 & 0.949 & 0.989
\lastline
\multirow{3}{1cm}{10} & & & & 30 & 0.891 & 0.945 & 0.989 \\
& 0.5 & 0.674 & 0.522 & 50 & 0.898 & 0.948 & 0.988 \\
& & & & 100 & 0.900 & 0.949 & 0.990
\lastline
\end{tabular}
\end{center}
\end{footnotesize}
\end{table}

\begin{table}[!p]
\begin{footnotesize}
\caption{Monte Carlo coverages for the proposed ELQB confidence intervals for $\theta_{20}$, at fixed $\theta_{10}$ and $\theta_{30}$. Normal distributions. Competitor approaches are ELP, ELB, IF, PEL,  AEL, and BTII.}
\label{tab:3}
\begin{center}
\begin{tabular}{@{}l r r r r r r r r r@{}}
\tblhead{
Sample size & $1 - \alpha$ & ELQB & ELP & ELB & IF & PEL & AEL & BTII}
\multicolumn{9}{l}{Scenario 1: $\mathcal{N}(0, 1)$, $\mathcal{N}(2.5, 1.1^2)$, $\mathcal{N}(3.69, 1.2^2)$, $\theta_{10} = \theta_{30} = 0.8$, $\theta_{20} = 0.5$} 
\lastline
\multirow{3}{2.5cm}{(30, 30, 30)} & 0.90 & 0.900 & 0.894 & 0.919 & 0.874 & 0.912 & 0.927 & 0.909 \\
& 0.95 & 0.949 & 0.949 & 0.963 & 0.931 & 0.956 & 0.963 & 0.951 \\
& 0.99 & 0.988 & 0.990 & 0.993 & 0.982 & 0.985 & 0.992 & 0.985
\lastline
\multirow{3}{2.5cm}{(50, 30, 30)} & 0.90 & 0.894 & 0.893 & 0.921 & 0.873 & 0.907 & 0.923 & 0.917 \\
& 0.95 & 0.942 & 0.950 & 0.964 & 0.932 & 0.954 & 0.962 & 0.953 \\
& 0.99 & 0.985 & 0.991 & 0.995 & 0.983 & 0.983 & 0.991 & 0.985
\lastline
\multirow{3}{2.5cm}{(50, 50, 50)} & 0.90 & 0.901 & 0.896 & 0.914 & 0.889 & 0.905 & 0.912 & 0.918 \\
& 0.95 & 0.947 & 0.951 & 0.960 & 0.940 & 0.943 & 0.949 & 0.954 \\
& 0.99 & 0.985 & 0.990 & 0.993 & 0.985 & 0.977 & 0.983 & 0.989
\lastline
\multirow{3}{2.5cm}{(100, 50, 50)} & 0.90 & 0.892 & 0.894 & 0.915 & 0.885 & 0.905 & 0.912 & 0.920 \\
& 0.95 & 0.942 & 0.948 & 0.957 & 0.936 & 0.939 & 0.949 & 0.956 \\
& 0.99 & 0.982 & 0.990 & 0.992 & 0.983 & 0.975 & 0.984 & 0.987
\lastline
\multirow{3}{2.5cm}{(100, 100, 50)} & 0.90 & 0.877 & 0.890 & 0.904 & 0.885 & 0.890 & 0.899 & 0.913 \\
& 0.95 & 0.928 & 0.944 & 0.949 & 0.939 & 0.937 & 0.941 & 0.951 \\
& 0.99 & 0.977 & 0.985 & 0.988 & 0.982 & 0.974 & 0.978 & 0.984 
\lastline
\multirow{3}{2.5cm}{(100, 100, 100)} & 0.90 & 0.897 & 0.897 & 0.913 & 0.892 & 0.893 & 0.902 & 0.921 \\
& 0.95 & 0.944 & 0.949 & 0.956 & 0.943 & 0.933 & 0.936 & 0.961 \\
& 0.99 & 0.986 & 0.992 & 0.991 & 0.990 & 0.975 & 0.977 & 0.990 
\lastline
\multicolumn{9}{l}{Scenario 2: $\mathcal{N}(0, 1)$, $\mathcal{N}(3.5, 1.1^2)$, $\mathcal{N}(5.5, 1.2^2)$, $\theta_{10} = \theta_{30} = 0.8$, $\theta_{20} = 0.8$} 
\lastline
\multirow{3}{2.5cm}{(30, 30, 30)} & 0.90 & 0.937 & 0.936 & 0.946 & 0.878 & 0.919 & 0.938 & 0.927 \\
& 0.95 & 0.973 & 0.973 & 0.978 & 0.932 & 0.951 & 0.957 & 0.964 \\
& 0.99 & 0.986 & 0.999 & 0.997 & 0.951 & 0.986 & 0.987 & 0.991
\lastline
\multirow{3}{2.5cm}{(50, 30, 30)} & 0.90 & 0.917 & 0.932 & 0.940 & 0.867 & 0.908 & 0.931 & 0.920 \\
& 0.95 & 0.961 & 0.975 & 0.974 & 0.928 & 0.947 & 0.951 & 0.955 \\
& 0.99 & 0.980 & 0.997 & 0.995 & 0.948 & 0.986 & 0.988 & 0.987
\lastline
\multirow{3}{2.5cm}{(50, 50, 50)} & 0.90 & 0.901 & 0.911 & 0.923 & 0.886 & 0.916 & 0.925 & 0.922 \\
& 0.95 & 0.949 & 0.960 & 0.962 & 0.933 & 0.951 & 0.963 & 0.960 \\
& 0.99 & 0.989 & 0.992 & 0.991 & 0.981 & 0.984 & 0.988 & 0.992 
\lastline
\multirow{3}{2.5cm}{(100, 50, 50)} & 0.90 & 0.899 & 0.909 & 0.923 & 0.887 & 0.910 & 0.921 & 0.925 \\
& 0.95 & 0.950 & 0.961 & 0.967 & 0.938 & 0.950 & 0.962 & 0.963 \\
& 0.99 & 0.988 & 0.993 & 0.992 & 0.981 & 0.983 & 0.988 & 0.991 
\lastline
\multirow{3}{2.5cm}{(100, 100, 50)} & 0.90 & 0.861 & 0.899 & 0.905 & 0.877 & 0.882 & 0.889 & 0.910 \\
& 0.95 & 0.921 & 0.953 & 0.949 & 0.934 & 0.934 & 0.939 & 0.948 \\
& 0.99 & 0.978 & 0.991 & 0.991 & 0.981 & 0.974 & 0.978 & 0.988 
\lastline
\multirow{3}{2.5cm}{(100, 100, 100)} & 0.90 & 0.883 & 0.895 & 0.907 & 0.879 & 0.888 & 0.892 & 0.911 \\
& 0.95 & 0.936 & 0.949 & 0.953 & 0.933 & 0.939 & 0.941 & 0.954 \\
& 0.99 & 0.982 & 0.991 & 0.990 & 0.980 & 0.974 & 0.976 & 0.988 
\lastline
\multicolumn{9}{l}{Scenario 3: $\mathcal{N}(0, 1)$, $\mathcal{N}(4, 1.2^2)$, $\mathcal{N}(8.189, 2^2)$, $\theta_{10} = \theta_{30} = 0.9$, $\theta_{20} = 0.9$} 
\lastline
\multirow{3}{2.5cm}{(30, 30, 30)} & 0.90 & 0.826 & 0.971 & 0.958 & 0.806 & 0.904 & 0.922 & 0.951 \\
& 0.95 & 0.839 & 0.987 & 0.978 & 0.820 & 0.952 & 0.956 & 0.976 \\
& 0.99 & 0.849 & 0.997 & 0.995 & 0.826 & 0.977 & 0.977 & 0.995 
\lastline
\multirow{3}{2.5cm}{(50, 30, 30)} & 0.90 & 0.821 & 0.971 & 0.958 & 0.803 & 0.896 & 0.918 & 0.949 \\
& 0.95 & 0.834 & 0.986 & 0.977 & 0.816 & 0.951 & 0.954 & 0.972 \\
& 0.99 & 0.845 & 0.998 & 0.996 & 0.824 & 0.970 & 0.974 & 0.993 
\lastline
\multirow{3}{2.5cm}{(50, 50, 50)} & 0.90 & 0.921 & 0.945 & 0.949 & 0.864 & 0.903 & 0.904 & 0.945 \\
& 0.95 & 0.937 & 0.980 & 0.976 & 0.888 & 0.944 & 0.947 & 0.974 \\
& 0.99 & 0.952 & 0.998 & 0.997 & 0.896 & 0.982 & 0.984 & 0.995 
\lastline
\multirow{3}{2.5cm}{(100, 50, 50)} & 0.90 & 0.915 & 0.941 & 0.948 & 0.859 & 0.893 & 0.900 & 0.938 \\
& 0.95 & 0.932 & 0.978 & 0.974 & 0.882 & 0.936 & 0.938 & 0.971 \\
& 0.99 & 0.949 & 0.997 & 0.993 & 0.891 & 0.979 & 0.982 & 0.991 
\lastline
\multirow{3}{2.5cm}{(100, 100, 50)} & 0.90 & 0.875 & 0.905 & 0.929 & 0.849 & 0.852 & 0.860 & 0.930 \\
& 0.95 & 0.933 & 0.956 & 0.964 & 0.897 & 0.909 & 0.928 & 0.960 \\
& 0.99 & 0.981 & 0.991 & 0.991 & 0.936 & 0.970 & 0.972 & 0.989 
\lastline
\multirow{3}{2.5cm}{(100, 100, 100)} & 0.90 & 0.893 & 0.917 & 0.934 & 0.882 & 0.895 & 0.902 & 0.935 \\
& 0.95 & 0.943 & 0.961 & 0.966 & 0.929 & 0.941 & 0.943 & 0.965 \\
& 0.99 & 0.986 & 0.993 & 0.993 & 0.965 & 0.975 & 0.976 & 0.993 
\lastline
\end{tabular}
\end{center}
\end{footnotesize}
\end{table}

\begin{table}[!p]
\begin{footnotesize}
\caption{Monte Carlo coverages for the proposed ELQB confidence intervals for $\theta_{20}$, at fixed $\theta_{10}$ and $\theta_{30}$. Mixed distributions. Competitor approaches are ELP, ELB, IF, PEL,  AEL, and BTII.}
\label{tab:4}
\begin{center}
\begin{tabular}{@{}l r r r r r r r r r@{}}
\tblhead{
Sample size & $1 - \alpha$ & ELQB & ELP & ELB & IF & PEL & AEL & BTII}
\lastline
\multicolumn{9}{l}{Scenario 4: $\mathcal{G}(6, 12)$, $\mathcal{LN}(1.5, 0.5)$, $\mathcal{W}(4, 6.6)$, $\theta_{10} = \theta_{30} = 0.8$, $\theta_{20} = 0.5$}
\lastline
\multirow{3}{2.5cm}{(30, 30, 30)} & 0.90 & 0.913 & 0.899 & 0.918 & 0.895 & 0.922 & 0.938 & 0.916 \\ 
& 0.95 & 0.960 & 0.948 & 0.962 & 0.943 & 0.966 & 0.973 & 0.954 \\ 
& 0.99 & 0.988 & 0.987 & 0.990 & 0.985 & 0.991 & 0.996 & 0.984 
\lastline
\multirow{3}{2.5cm}{(50, 30, 30)} & 0.90 & 0.912 & 0.898 & 0.921 & 0.896 & 0.927 & 0.943 & 0.918 \\ 
& 0.95 & 0.957 & 0.945 & 0.959 & 0.942 & 0.968 & 0.973 & 0.954 \\ 
& 0.99 & 0.988 & 0.989 & 0.992 & 0.986 & 0.990 & 0.995 & 0.985 
\lastline
\multirow{3}{2.5cm}{(50, 50, 50)} & 0.90 & 0.900 & 0.883 & 0.905 & 0.881 & 0.913 & 0.922 & 0.913 \\ 
& 0.95 & 0.947 & 0.939 & 0.954 & 0.934 & 0.953 & 0.961 & 0.951 \\ 
& 0.99 & 0.986 & 0.985 & 0.989 & 0.983 & 0.984 & 0.990 & 0.984 
\lastline
\multirow{3}{2.5cm}{(100, 50, 50)} & 0.90 & 0.893 & 0.887 & 0.909 & 0.884 & 0.915 & 0.924 & 0.917 \\ 
& 0.95 & 0.945 & 0.941 & 0.956 & 0.940 & 0.952 & 0.962 & 0.955 \\ 
& 0.99 & 0.986 & 0.985 & 0.990 & 0.984 & 0.984 & 0.992 & 0.985 
\lastline
\multirow{3}{2.5cm}{(100, 100, 50)} & 0.90 & 0.879 & 0.877 & 0.898 & 0.876 & 0.902 & 0.911 & 0.908 \\ 
& 0.95 & 0.928 & 0.933 & 0.944 & 0.933 & 0.943 & 0.946 & 0.948 \\ 
& 0.99 & 0.978 & 0.984 & 0.985 & 0.984 & 0.979 & 0.983 & 0.985 
\lastline
\multirow{3}{2.5cm}{(100, 100, 100)} & 0.90 & 0.892 & 0.888 & 0.906 & 0.888 & 0.898 & 0.910 & 0.912 \\ 
& 0.95 & 0.942 & 0.940 & 0.951 & 0.939 & 0.941 & 0.943 & 0.955 \\ 
& 0.99 & 0.986 & 0.988 & 0.990 & 0.987 & 0.983 & 0.986 & 0.990 
\lastline
\multicolumn{9}{l}{Scenario 5: $\mathcal{G}(6, 12)$, $\mathcal{LN}(1.5, 0.5)$, $\mathcal{W}(4, 10)$, $\theta_{10} = \theta_{30} = 0.8$, $\theta_{20} = 0.8$} 
\lastline
\multirow{3}{2.5cm}{(30, 30, 30)} & 0.90 & 0.939 & 0.926 & 0.941 & 0.907 & 0.923 & 0.946 & 0.931 \\ 
& 0.95 & 0.979 & 0.978 & 0.973 & 0.947 & 0.957 & 0.963 & 0.963 \\ 
& 0.99 & 0.993 & 0.998 & 0.997 & 0.956 & 0.992 & 0.993 & 0.991 
\lastline
\multirow{3}{2.5cm}{(50, 30, 30)} & 0.90 & 0.931 & 0.928 & 0.939 & 0.901 & 0.930 & 0.952 & 0.928 \\ 
& 0.95 & 0.973 & 0.976 & 0.974 & 0.948 & 0.962 & 0.963 & 0.964 \\ 
& 0.99 & 0.992 & 0.998 & 0.996 & 0.959 & 0.991 & 0.992 & 0.991 
\lastline
\multirow{3}{2.5cm}{(50, 50, 50)} & 0.90 & 0.912 & 0.921 & 0.928 & 0.904 & 0.920 & 0.934 & 0.931 \\ 
& 0.95 & 0.959 & 0.964 & 0.967 & 0.951 & 0.954 & 0.968 & 0.964 \\ 
& 0.99 & 0.993 & 0.996 & 0.995 & 0.988 & 0.988 & 0.992 & 0.993 
\lastline
\multirow{3}{2.5cm}{(100, 50, 50)} & 0.90 & 0.904 & 0.920 & 0.922 & 0.902 & 0.915 & 0.928 & 0.922 \\ 
& 0.95 & 0.948 & 0.960 & 0.962 & 0.951 & 0.950 & 0.964 & 0.959 \\ 
& 0.99 & 0.991 & 0.994 & 0.993 & 0.984 & 0.987 & 0.992 & 0.991 
\lastline
\multirow{3}{2.5cm}{(100, 100, 50)} & 0.90 & 0.897 & 0.915 & 0.912 & 0.905 & 0.905 & 0.914 & 0.923 \\ 
& 0.95 & 0.944 & 0.958 & 0.961 & 0.948 & 0.953 & 0.956 & 0.962 \\ 
& 0.99 & 0.986 & 0.993 & 0.993 & 0.987 & 0.983 & 0.985 & 0.990 
\lastline
\multirow{3}{2.5cm}{(100, 100, 100)} & 0.90 & 0.902 & 0.917 & 0.917 & 0.909 & 0.900 & 0.901 & 0.924 \\ 
& 0.95 & 0.955 & 0.962 & 0.959 & 0.957 & 0.951 & 0.953 & 0.963 \\ 
& 0.99 & 0.991 & 0.993 & 0.994 & 0.990 & 0.986 & 0.987 & 0.993  
\lastline
\multicolumn{9}{l}{Scenario 6: $\mathcal{G}(6, 12)$, $\mathcal{LN}(1.5, 0.5)$, $\mathcal{W}(4, 12.4)$, $\theta_{10} = \theta_{30} = 0.8$, $\theta_{20} = 0.9$} 
\lastline
\multirow{3}{2.5cm}{(30, 30, 30)} & 0.90 & 0.891 & 0.964 & 0.960 & 0.804 & 0.914 & 0.940 & 0.918 \\ 
& 0.95 & 0.908 & 0.985 & 0.978 & 0.822 & 0.955 & 0.965 & 0.952 \\ 
& 0.99 & 0.919 & 0.998 & 0.996 & 0.841 & 0.969 & 0.976 & 0.982 
\lastline
\multirow{3}{2.5cm}{(50, 30, 30)} & 0.90 & 0.903 & 0.974 & 0.968 & 0.827 & 0.924 & 0.952 & 0.933 \\ 
& 0.95 & 0.917 & 0.991 & 0.984 & 0.837 & 0.963 & 0.974 & 0.960 \\ 
& 0.99 & 0.927 & 0.998 & 0.998 & 0.852 & 0.976 & 0.981 & 0.989  
\lastline
\multirow{3}{2.5cm}{(50, 50, 50)} & 0.90 & 0.948 & 0.943 & 0.943 & 0.898 & 0.916 & 0.928 & 0.927 \\ 
& 0.95 & 0.972 & 0.975 & 0.971 & 0.914 & 0.951 & 0.951 & 0.959 \\ 
& 0.99 & 0.985 & 0.998 & 0.996 & 0.919 & 0.991 & 0.990 & 0.989 
\lastline
\multirow{3}{2.5cm}{(100, 50, 50)} & 0.90 & 0.943 & 0.939 & 0.939 & 0.889 & 0.920 & 0.941 & 0.924 \\ 
& 0.95 & 0.970 & 0.974 & 0.971 & 0.906 & 0.954 & 0.954 & 0.960 \\ 
& 0.99 & 0.984 & 0.998 & 0.995 & 0.912 & 0.990 & 0.990 & 0.990 
\lastline
\multirow{3}{2.5cm}{(100, 100, 50)} & 0.90 & 0.904 & 0.925 & 0.927 & 0.914 & 0.913 & 0.921 & 0.934 \\ 
& 0.95 & 0.950 & 0.968 & 0.965 & 0.951 & 0.955 & 0.956 & 0.968 \\ 
& 0.99 & 0.991 & 0.995 & 0.993 & 0.976 & 0.984 & 0.985 & 0.993 
\lastline
\multirow{3}{2.5cm}{(100, 100, 100)} & 0.90 & 0.910 & 0.928 & 0.926 & 0.915 & 0.917 & 0.918 & 0.931 \\ 
& 0.95 & 0.955 & 0.966 & 0.966 & 0.958 & 0.958 & 0.959 & 0.967 \\ 
& 0.99 & 0.990 & 0.996 & 0.994 & 0.979 & 0.985 & 0.985 & 0.992 
\lastline
\end{tabular}
\end{center}
\end{footnotesize}
\end{table}

\begin{table}[!p]
\begin{footnotesize}
\caption{Monte Carlo coverages for the proposed ELQB confidence intervals for $\theta_{20}$, at fixed $\theta_{10}$ and $\theta_{30}$. Beta distributions. Competitor approaches are ELP, ELB, IF, PEL,  AEL, and BTII.}
\label{tab:5}
\begin{center}
\begin{tabular}{@{}l r r r r r r r r r@{}}
\tblhead{
Sample size & $1 - \alpha$ & ELQB & ELP & ELB & IF & PEL & AEL & BTII}
\multicolumn{9}{l}{Scenario 7: $\mathcal{B}(1, 6)$, $\mathcal{B}(6, 6)$, $\mathcal{B}(9.6, 6)$, $\theta_{10} = \theta_{30} = 0.8$, $\theta_{20} = 0.5$}
\lastline
\multirow{3}{2.5cm}{(30, 30, 30)} & 0.90 & 0.915 & 0.897 & 0.926 & 0.887 & 0.918 & 0.934 & 0.925 \\ 
& 0.95 & 0.960 & 0.950 & 0.969 & 0.938 & 0.961 & 0.969 & 0.961 \\ 
& 0.99 & 0.990 & 0.991 & 0.994 & 0.985 & 0.990 & 0.993 & 0.988 
\lastline
\multirow{3}{2.5cm}{(50, 30, 30)} & 0.90 & 0.903 & 0.890 & 0.918 & 0.878 & 0.915 & 0.932 & 0.916 \\ 
& 0.95 & 0.948 & 0.940 & 0.958 & 0.930 & 0.959 & 0.968 & 0.951 \\ 
& 0.99 & 0.988 & 0.989 & 0.994 & 0.981 & 0.989 & 0.994 & 0.984
\lastline
\multirow{3}{2.5cm}{(50, 50, 50)} & 0.90 & 0.905 & 0.890 & 0.913 & 0.883 & 0.917 & 0.928 & 0.918 \\ 
& 0.95 & 0.947 & 0.939 & 0.954 & 0.936 & 0.958 & 0.966 & 0.957 \\ 
& 0.99 & 0.986 & 0.987 & 0.992 & 0.985 & 0.989 & 0.992 & 0.988 
\lastline
\multirow{3}{2.5cm}{(100, 50, 50)} & 0.90 & 0.897 & 0.888 & 0.906 & 0.883 & 0.913 & 0.923 & 0.912 \\ 
& 0.95 & 0.944 & 0.939 & 0.954 & 0.936 & 0.958 & 0.966 & 0.954 \\ 
& 0.99 & 0.985 & 0.985 & 0.991 & 0.982 & 0.990 & 0.992 & 0.985 
\lastline
\multirow{3}{2.5cm}{(100, 100, 50)} & 0.90 & 0.882 & 0.889 & 0.903 & 0.888 & 0.905 & 0.913 & 0.910 \\ 
& 0.95 & 0.934 & 0.937 & 0.949 & 0.937 & 0.955 & 0.961 & 0.950 \\ 
& 0.99 & 0.981 & 0.984 & 0.986 & 0.983 & 0.990 & 0.992 & 0.986 
\lastline
\multirow{3}{2.5cm}{(100, 100, 100)} & 0.90 & 0.898 & 0.890 & 0.908 & 0.887 & 0.912 & 0.920 & 0.915 \\ 
& 0.95 & 0.946 & 0.941 & 0.951 & 0.940 & 0.958 & 0.962 & 0.954 \\ 
& 0.99 & 0.985 & 0.985 & 0.989 & 0.983 & 0.991 & 0.992 & 0.987 
\lastline
\multicolumn{9}{l}{Scenario 8: $\mathcal{B}(1, 6)$, $\mathcal{B}(9, 6)$, $\mathcal{B}(20.4, 6)$, $\theta_{10} = \theta_{30} = 0.8$, $\theta_{20} = 0.8$}
\lastline
\multirow{3}{2.5cm}{(30, 30, 30)} & 0.90 & 0.931 & 0.904 & 0.932 & 0.874 & 0.913 & 0.934 & 0.922 \\ 
& 0.95 & 0.974 & 0.961 & 0.973 & 0.933 & 0.956 & 0.967 & 0.960 \\ 
& 0.99 & 0.990 & 0.994 & 0.994 & 0.952 & 0.991 & 0.992 & 0.987 
\lastline
\multirow{3}{2.5cm}{(50, 30, 30)} & 0.90 & 0.935 & 0.907 & 0.935 & 0.876 & 0.917 & 0.937 & 0.924 \\ 
& 0.95 & 0.975 & 0.963 & 0.971 & 0.933 & 0.958 & 0.966 & 0.957 \\ 
& 0.99 & 0.993 & 0.995 & 0.994 & 0.952 & 0.990 & 0.992 & 0.988 
\lastline
\multirow{3}{2.5cm}{(50, 50, 50)} & 0.90 & 0.903 & 0.893 & 0.916 & 0.880 & 0.912 & 0.925 & 0.921 \\ 
& 0.95 & 0.958 & 0.946 & 0.963 & 0.937 & 0.955 & 0.966 & 0.959 \\ 
& 0.99 & 0.992 & 0.993 & 0.992 & 0.986 & 0.992 & 0.993 & 0.989 
\lastline
\multirow{3}{2.5cm}{(100, 50, 50)} & 0.90 & 0.906 & 0.894 & 0.918 & 0.884 & 0.914 & 0.925 & 0.922 \\ 
& 0.95 & 0.955 & 0.948 & 0.963 & 0.936 & 0.957 & 0.964 & 0.956 \\ 
& 0.99 & 0.990 & 0.989 & 0.990 & 0.984 & 0.990 & 0.993 & 0.987 
\lastline
\multirow{3}{2.5cm}{(100, 100, 50)} & 0.90 & 0.897 & 0.898 & 0.911 & 0.891 & 0.909 & 0.919 & 0.921 \\ 
& 0.95 & 0.940 & 0.945 & 0.956 & 0.940 & 0.958 & 0.964 & 0.961 \\ 
& 0.99 & 0.985 & 0.988 & 0.992 & 0.983 & 0.991 & 0.993 & 0.991 
\lastline
\multirow{3}{2.5cm}{(100, 100, 100)} & 0.90 & 0.904 & 0.893 & 0.910 & 0.893 & 0.906 & 0.914 & 0.912 \\ 
& 0.95 & 0.948 & 0.944 & 0.957 & 0.942 & 0.955 & 0.960 & 0.957 \\ 
& 0.99 & 0.987 & 0.990 & 0.990 & 0.986 & 0.990 & 0.993 & 0.990 
\lastline
\multicolumn{9}{l}{Scenario 9: $\mathcal{B}(1, 6)$, $\mathcal{B}(6, 6)$, $\mathcal{B}(20.4, 6)$, $\theta_{10} = \theta_{30} = 0.8$, $\theta_{20} = 0.9$} 
\lastline
\multirow{3}{2.5cm}{(30, 30, 30)} & 0.90 & 0.912 & 0.960 & 0.964 & 0.830 & 0.937 & 0.953 & 0.952 \\ 
& 0.95 & 0.925 & 0.981 & 0.982 & 0.851 & 0.967 & 0.974 & 0.977 \\ 
& 0.99 & 0.935 & 0.996 & 0.997 & 0.874 & 0.985 & 0.986 & 0.993 
\lastline
\multirow{3}{2.5cm}{(50, 30, 30)} & 0.90 & 0.910 & 0.962 & 0.968 & 0.815 & 0.938 & 0.952 & 0.944 \\ 
& 0.95 & 0.924 & 0.981 & 0.983 & 0.831 & 0.967 & 0.971 & 0.968 \\ 
& 0.99 & 0.933 & 0.995 & 0.997 & 0.858 & 0.984 & 0.986 & 0.989 
\lastline
\multirow{3}{2.5cm}{(50, 50, 50)} & 0.90 & 0.954 & 0.943 & 0.956 & 0.906 & 0.929 & 0.940 & 0.942 \\ 
& 0.95 & 0.974 & 0.978 & 0.979 & 0.923 & 0.963 & 0.967 & 0.969 \\ 
& 0.99 & 0.986 & 0.997 & 0.996 & 0.934 & 0.993 & 0.996 & 0.994 
\lastline
\multirow{3}{2.5cm}{(100, 50, 50)} & 0.90 & 0.949 & 0.939 & 0.954 & 0.889 & 0.926 & 0.936 & 0.935 \\ 
& 0.95 & 0.972 & 0.978 & 0.979 & 0.905 & 0.961 & 0.966 & 0.968 \\ 
& 0.99 & 0.983 & 0.997 & 0.997 & 0.916 & 0.995 & 0.996 & 0.992 
\lastline
\multirow{3}{2.5cm}{(100, 100, 50)} & 0.90 & 0.921 & 0.919 & 0.942 & 0.907 & 0.929 & 0.934 & 0.942 \\ 
& 0.95 & 0.966 & 0.964 & 0.976 & 0.946 & 0.966 & 0.971 & 0.974 \\ 
& 0.99 & 0.994 & 0.997 & 0.996 & 0.974 & 0.994 & 0.995 & 0.995 
\lastline
\multirow{3}{2.5cm}{(100, 100, 100)} & 0.90 & 0.916 & 0.916 & 0.934 & 0.914 & 0.914 & 0.920 & 0.933 \\ 
& 0.95 & 0.957 & 0.962 & 0.965 & 0.949 & 0.960 & 0.964 & 0.966 \\ 
& 0.99 & 0.993 & 0.993 & 0.994 & 0.981 & 0.990 & 0.993 & 0.991 
\lastline
\end{tabular}
\end{center}
\end{footnotesize}
\end{table}

\begin{table}[!p]
\begin{footnotesize}
\caption{Monte Carlo coverages for the proposed ELQB confidence intervals for $\theta_{20}$, at fixed $\theta_{10}$ and $\theta_{30}$. Mixture distributions. Competitor approaches are ELP, ELB, IF, PEL,  AEL, and BTII.}
\label{tab:6}
\begin{center}
\begin{tabular}{@{}l r r r r r r r r r@{}}
\tblhead{
Sample size & $1 - \alpha$ & ELQB & ELP & ELB & IF & PEL & AEL & BTII}
\multicolumn{9}{l}{Scenario 10: $0.5N(-1, 1) + 0.5N(2, 1)$, $0.5N(1, 1) + 0.5N(4, 1.5)$, $0.5N(3, 1.5) + 0.5N(6, 1)$}\\
\multicolumn{9}{l}{$\theta_{10} = 0.5, \theta_{30} = 0.522$, $\theta_{20} = 0.674$}
\lastline
\multirow{3}{2.5cm}{(30, 30, 30)} & 0.90 & 0.897 & 0.851 & 0.914 & 0.821 & 0.911 & 0.914 & 0.921 \\ 
& 0.95 & 0.945 & 0.921 & 0.957 & 0.891 & 0.932 & 0.944 & 0.956 \\ 
& 0.99 & 0.985 & 0.975 & 0.987 & 0.962 & 0.976 & 0.973 & 0.986 
\lastline
\multirow{3}{2.5cm}{(50, 30, 30)} & 0.90 & 0.909 & 0.859 & 0.927 & 0.843 & 0.899 & 0.915 & 0.928 \\ 
& 0.95 & 0.956 & 0.928 & 0.965 & 0.902 & 0.919 & 0.949 & 0.961 \\ 
& 0.99 & 0.991 & 0.981 & 0.992 & 0.970 & 0.975 & 0.982 & 0.990 
\lastline
\multirow{3}{2.5cm}{(50, 50, 50)} & 0.90 & 0.888 & 0.840 & 0.909 & 0.825 & 0.864 & 0.885 & 0.921 \\ 
& 0.95 & 0.938 & 0.907 & 0.955 & 0.895 & 0.902 & 0.924 & 0.956 \\ 
& 0.99 & 0.983 & 0.973 & 0.991 & 0.964 & 0.956 & 0.964 & 0.988 
\lastline
\multirow{3}{2.5cm}{(100, 50, 50)} & 0.90 & 0.888 & 0.842 & 0.910 & 0.829 & 0.847 & 0.874 & 0.918 \\ 
& 0.95 & 0.938 & 0.907 & 0.955 & 0.898 & 0.897 & 0.920 & 0.957 \\ 
& 0.99 & 0.983 & 0.973 & 0.991 & 0.966 & 0.956 & 0.963 & 0.989 
\lastline
\multirow{3}{2.5cm}{(100, 100, 50)} & 0.90 & 0.883 & 0.830 & 0.899 & 0.819 & 0.862 & 0.867 & 0.916 \\ 
& 0.95 & 0.936 & 0.898 & 0.952 & 0.884 & 0.898 & 0.905 & 0.956 \\ 
& 0.99 & 0.982 & 0.975 & 0.991 & 0.965 & 0.946 & 0.953 & 0.989 
\lastline
\multirow{3}{2.5cm}{(100, 100, 100)} & 0.90 & 0.890 & 0.834 & 0.904 & 0.827 & 0.851 & 0.851 & 0.918 \\ 
& 0.95 & 0.939 & 0.905 & 0.954 & 0.899 & 0.891 & 0.897 & 0.960 \\ 
& 0.99 & 0.984 & 0.977 & 0.991 & 0.968 & 0.946 & 0.954 & 0.990 
\lastline
\end{tabular}
\end{center}
\end{footnotesize}
\end{table}

\begin{table}[!p]
\begin{footnotesize}
\caption{Monte Carlo coverages for the proposed ELQB confidence intervals for the VUS. Normal distributions. ELU and JEL are competitor approaches.}
\label{tab:vus:1}
\begin{center}
\begin{tabular}{@{}l r r r r r r r r r@{}}
\tblhead{
& \multicolumn{3}{c}{$1 - \alpha = 0.90$} & \multicolumn{3}{c}{$1 - \alpha = 0.95$} & \multicolumn{3}{c}{$1 - \alpha = 0.99$} \\
Sample size & ELQB & ELU & JEL & ELQB & ELU & JEL & ELQB & ELU & JEL}
\multicolumn{10}{l}{Scenario 1: $\mathcal{N}(0, 1)$, $\mathcal{N}(2.5, 1.1^2)$, $\mathcal{N}(3.69, 1.2^2)$, $\gamma_0 = 0.722$} \\
(15, 15, 15) & 0.887 & 0.887 & 0.919 & 0.935 & 0.928 & 0.955 & 0.982 & 0.968 & 0.988 \\ 
(30, 30, 30) & 0.896 & 0.907 & 0.910 & 0.945 & 0.953 & 0.954 & 0.985 & 0.986 & 0.992 \\ 
(50, 30, 30) & 0.898 & 0.906 & 0.909 & 0.941 & 0.953 & 0.954 & 0.983 & 0.986 & 0.988 \\ 
(50, 50, 50) & 0.899 & 0.908 & 0.909 & 0.946 & 0.954 & 0.958 & 0.987 & 0.992 & 0.992 \\ 
(75, 75, 75) & 0.903 & 0.915 & 0.915 & 0.950 & 0.958 & 0.960 & 0.990 & 0.991 & 0.993 \\ 
(100, 100, 100) & 0.891 & 0.903 & 0.901 & 0.942 & 0.949 & 0.951 & 0.987 & 0.988 & 0.990 
\lastline
\multicolumn{10}{l}{Scenario 2: $\mathcal{N}(0, 1)$, $\mathcal{N}(3.5, 1.1^2)$, $\mathcal{N}(5.5, 1.2^2)$, $\gamma_0 = 0.881$} \\
(15, 15, 15) & 0.889 & 0.852 & 0.885 & 0.946 & 0.899 & 0.923 & 0.979 & 0.943 & 0.960 \\ 
(30, 30, 30) & 0.884 & 0.884 & 0.895 & 0.939 & 0.936 & 0.946 & 0.981 & 0.980 & 0.980 \\ 
(50, 30, 30) & 0.891 & 0.887 & 0.897 & 0.933 & 0.937 & 0.939 & 0.978 & 0.979 & 0.981 \\ 
(50, 50, 50) & 0.888 & 0.900 & 0.903 & 0.943 & 0.950 & 0.952 & 0.983 & 0.988 & 0.988 \\ 
(75, 75, 75) & 0.887 & 0.900 & 0.902 & 0.941 & 0.951 & 0.951 & 0.987 & 0.989 & 0.990 \\ 
(100, 100, 100) & 0.890 & 0.904 & 0.902 & 0.943 & 0.954 & 0.952 & 0.984 & 0.991 & 0.991 
\lastline
\multicolumn{10}{l}{Scenario 3: $\mathcal{N}(0, 1)$, $\mathcal{N}(4, 1.2^2)$, $\mathcal{N}(8.189, 2^2)$, $\gamma_0 = 0.959$} \\
(15, 15, 15) & 0.889 & 0.731 & 0.767 & 0.909 & 0.785 & 0.791 & 0.919 & 0.829 & 0.854 \\ 
(30, 30, 30) & 0.836 & 0.810 & 0.826 & 0.898 & 0.862 & 0.877 & 0.960 & 0.920 & 0.929 \\ 
(50, 30, 30) & 0.855 & 0.822 & 0.840 & 0.905 & 0.877 & 0.887 & 0.956 & 0.929 & 0.935 \\ 
(50, 50, 50) & 0.859 & 0.851 & 0.860 & 0.912 & 0.901 & 0.910 & 0.965 & 0.959 & 0.961 \\ 
(75, 75, 75) & 0.874 & 0.878 & 0.883 & 0.922 & 0.929 & 0.931 & 0.973 & 0.975 & 0.975 \\ 
(100, 100, 100) & 0.873 & 0.879 & 0.881 & 0.925 & 0.930 & 0.936 & 0.974 & 0.977 & 0.979 
\lastline
\end{tabular}
\end{center}
\end{footnotesize}
\end{table}

\begin{table}[!p]
\begin{footnotesize}
\caption{Monte Carlo coverages for the proposed ELQB confidence intervals for the VUS. Mixed distributions. ELU and JEL are competitor approaches.}
\label{tab:vus:2}
\begin{center}
\begin{tabular}{@{}l r r r r r r r r r@{}}
\tblhead{
& \multicolumn{3}{c}{$1 - \alpha = 0.90$} & \multicolumn{3}{c}{$1 - \alpha = 0.95$} & \multicolumn{3}{c}{$1 - \alpha = 0.99$} \\
Sample size & ELQB & ELU & JEL & ELQB & ELU & JEL & ELQB & ELU & JEL}
\multicolumn{10}{l}{Scenario 4: $\mathcal{G}(6, 12)$, $\mathcal{LN}(1.5, 0.5)$, $\mathcal{W}(4, 6.6)$, $\gamma_0 = 0.669$} \\
(15, 15, 15) & 0.916 & 0.933 & 0.937 & 0.957 & 0.967 & 0.968 & 0.988 & 0.987 & 0.989 \\ 
(30, 30, 30) & 0.897 & 0.936 & 0.940 & 0.950 & 0.977 & 0.974 & 0.990 & 0.996 & 0.995 \\ 
(50, 30, 30) & 0.895 & 0.938 & 0.938 & 0.948 & 0.978 & 0.975 & 0.990 & 0.997 & 0.995 \\ 
(50, 50, 50) & 0.888 & 0.932 & 0.933 & 0.942 & 0.974 & 0.975 & 0.988 & 0.996 & 0.995 \\ 
(75, 75, 75) & 0.870 & 0.907 & 0.910 & 0.928 & 0.959 & 0.960 & 0.983 & 0.995 & 0.996 \\ 
(100, 100, 100) & 0.879 & 0.906 & 0.910 & 0.933 & 0.957 & 0.961 & 0.980 & 0.992 & 0.993
\lastline
\multicolumn{10}{l}{Scenario 5: $\mathcal{G}(6, 12)$, $\mathcal{LN}(1.5, 0.5)$, $\mathcal{W}(4, 10)$, $\gamma_0 = 0.868$} \\
(15, 15, 15) & 0.867 & 0.833 & 0.861 & 0.915 & 0.882 & 0.906 & 0.967 & 0.926 & 0.947 \\ 
(30, 30, 30) & 0.883 & 0.883 & 0.895 & 0.936 & 0.936 & 0.943 & 0.980 & 0.980 & 0.982 \\ 
(50, 30, 30) & 0.878 & 0.879 & 0.888 & 0.933 & 0.931 & 0.935 & 0.978 & 0.979 & 0.978 \\ 
(50, 50, 50) & 0.886 & 0.897 & 0.899 & 0.935 & 0.944 & 0.948 & 0.981 & 0.985 & 0.987 \\ 
(75, 75, 75) & 0.892 & 0.901 & 0.902 & 0.938 & 0.950 & 0.948 & 0.984 & 0.989 & 0.989 \\ 
(100, 100, 100) & 0.888 & 0.902 & 0.899 & 0.937 & 0.950 & 0.949 & 0.984 & 0.991 & 0.990
\lastline
\multicolumn{10}{l}{Scenario 6: $\mathcal{G}(6, 12)$, $\mathcal{LN}(1.5, 0.5)$, $\mathcal{W}(4, 12.4)$, $\gamma_0 = 0.927$} \\
(15, 15, 15) & 0.898 & 0.765 & 0.824 & 0.932 & 0.821 & 0.868 & 0.961 & 0.873 & 0.916 \\ 
(30, 30, 30) & 0.861 & 0.845 & 0.868 & 0.912 & 0.902 & 0.909 & 0.966 & 0.954 & 0.961 \\ 
(50, 30, 30) & 0.868 & 0.849 & 0.864 & 0.916 & 0.903 & 0.913 & 0.969 & 0.958 & 0.961 \\ 
(50, 50, 50) & 0.870 & 0.872 & 0.877 & 0.919 & 0.923 & 0.924 & 0.971 & 0.970 & 0.971 \\ 
(75, 75, 75) & 0.883 & 0.893 & 0.892 & 0.933 & 0.942 & 0.941 & 0.980 & 0.982 & 0.985 \\ 
(100, 100, 100) & 0.891 & 0.905 & 0.906 & 0.944 & 0.955 & 0.953 & 0.984 & 0.988 & 0.989 
\lastline
\end{tabular}
\end{center}
\end{footnotesize}
\end{table}

\begin{table}[!p]
\begin{footnotesize}
\caption{Monte Carlo coverages for the proposed ELQB confidence intervals for the VUS. Beta distributions. ELU and JEL are competitor approaches.}
\label{tab:vus:3}
\begin{center}
\begin{tabular}{@{}l r r r r r r r r r@{}}
\tblhead{
& \multicolumn{3}{c}{$1 - \alpha = 0.90$} & \multicolumn{3}{c}{$1 - \alpha = 0.95$} & \multicolumn{3}{c}{$1 - \alpha = 0.99$} \\
Sample size & ELQB & ELU & JEL & ELQB & ELU & JEL & ELQB & ELU & JEL}
\lastline
\multicolumn{10}{l}{Scenario 7: $\mathcal{B}(1, 6)$, $\mathcal{B}(6, 6)$, $\mathcal{B}(9.6, 6)$, $\gamma_0 = 0.698$} \\
(15, 15, 15) & 0.867 & 0.890 & 0.915 & 0.923 & 0.936 & 0.962 & 0.984 & 0.979 & 0.992 \\ 
(30, 30, 30) & 0.885 & 0.902 & 0.906 & 0.937 & 0.951 & 0.954 & 0.981 & 0.990 & 0.993 \\ 
(50, 30, 30) & 0.894 & 0.914 & 0.912 & 0.941 & 0.956 & 0.956 & 0.981 & 0.991 & 0.991 \\ 
(50, 50, 50) & 0.896 & 0.911 & 0.907 & 0.945 & 0.958 & 0.958 & 0.987 & 0.992 & 0.991 \\ 
(75, 75, 75) & 0.893 & 0.902 & 0.902 & 0.944 & 0.951 & 0.951 & 0.982 & 0.988 & 0.988 \\ 
(100, 100, 100) & 0.897 & 0.910 & 0.907 & 0.947 & 0.955 & 0.954 & 0.989 & 0.991 & 0.991 
\lastline
\multicolumn{10}{l}{Scenario 8: $\mathcal{B}(1, 6)$, $\mathcal{B}(9, 6)$, $\mathcal{B}(20.4, 6)$, $\gamma_0 = 0.869$} \\
(15, 15, 15) & 0.877 & 0.840 & 0.875 & 0.922 & 0.888 & 0.921 & 0.977 & 0.937 & 0.959 \\ 
(30, 30, 30) & 0.891 & 0.892 & 0.906 & 0.941 & 0.943 & 0.945 & 0.983 & 0.983 & 0.985 \\ 
(50, 30, 30) & 0.880 & 0.887 & 0.889 & 0.935 & 0.938 & 0.943 & 0.979 & 0.981 & 0.982 \\ 
(50, 50, 50) & 0.891 & 0.902 & 0.903 & 0.941 & 0.955 & 0.953 & 0.985 & 0.988 & 0.991 \\ 
(75, 75, 75) & 0.891 & 0.903 & 0.900 & 0.943 & 0.951 & 0.949 & 0.984 & 0.990 & 0.989 \\ 
(100, 100, 100) & 0.900 & 0.913 & 0.911 & 0.949 & 0.961 & 0.957 & 0.989 & 0.993 & 0.993 
\lastline
\multicolumn{10}{l}{Scenario 9: $\mathcal{B}(1, 6)$, $\mathcal{B}(6, 6)$, $\mathcal{B}(20.4, 6)$, $\gamma_0 = 0.917$} \\
(15, 15, 15) & 0.881 & 0.800 & 0.849 & 0.941 & 0.849 & 0.892 & 0.973 & 0.907 & 0.941 \\ 
(30, 30, 30) & 0.875 & 0.856 & 0.879 & 0.931 & 0.913 & 0.928 & 0.975 & 0.962 & 0.973 \\ 
(50, 30, 30) & 0.889 & 0.874 & 0.893 & 0.939 & 0.926 & 0.943 & 0.985 & 0.975 & 0.982 \\ 
(50, 50, 50) & 0.889 & 0.882 & 0.897 & 0.937 & 0.937 & 0.943 & 0.982 & 0.976 & 0.983 \\ 
(75, 75, 75) & 0.888 & 0.892 & 0.891 & 0.938 & 0.939 & 0.946 & 0.983 & 0.982 & 0.988 \\ 
(100, 100, 100) & 0.882 & 0.897 & 0.897 & 0.939 & 0.944 & 0.947 & 0.985 & 0.981 & 0.990 
\lastline
\end{tabular}
\end{center}
\end{footnotesize}
\end{table}

\begin{table}[!p]
\begin{footnotesize}
\caption{Monte Carlo coverages for the proposed ELQB confidence intervals for the VUS. Mixture distributions. ELU and JEL are competitor approaches.}
\label{tab:vus:4}
\begin{center}
\begin{tabular}{@{}l r r r r r r r r r@{}}
\tblhead{
& \multicolumn{3}{c}{$1 - \alpha = 0.90$} & \multicolumn{3}{c}{$1 - \alpha = 0.95$} & \multicolumn{3}{c}{$1 - \alpha = 0.99$} \\
Sample size & ELQB & ELU & JEL & ELQB & ELU & JEL & ELQB & ELU & JEL}
\multicolumn{10}{l}{Scenario 10: $0.5N(-1, 1) + 0.5N(2, 1)$, $0.5N(1, 1) + 0.5N(4, 1.5)$, $0.5N(3, 1.5) + 0.5N(6, 1)$} \\
\multicolumn{10}{l}{$\gamma_0 = 0.544$} \\
(15, 15, 15) & 0.900 & 0.877 & 0.921 & 0.948 & 0.916 & 0.965 & 0.990 & 0.947 & 0.994 \\ 
(30, 30, 30) & 0.897 & 0.898 & 0.907 & 0.949 & 0.941 & 0.956 & 0.989 & 0.978 & 0.992 \\ 
(50, 30, 30) & 0.896 & 0.903 & 0.908 & 0.944 & 0.944 & 0.956 & 0.988 & 0.980 & 0.993 \\ 
(50, 50, 50) & 0.901 & 0.911 & 0.911 & 0.951 & 0.955 & 0.960 & 0.989 & 0.988 & 0.993 \\ 
(75, 75, 75) & 0.895 & 0.904 & 0.904 & 0.947 & 0.951 & 0.953 & 0.985 & 0.985 & 0.990 \\ 
(100, 100, 100) & 0.896 & 0.910 & 0.906 & 0.947 & 0.953 & 0.955 & 0.988 & 0.987 & 0.990 
\lastline
\end{tabular}
\end{center}
\end{footnotesize}
\end{table}

\begin{table}[htpb]
\begin{footnotesize}
\caption{Monte Carlo coverages for the proposed ELQB confidence regions for the pair $(\theta_{20}, \theta_{30})$, at fixed $\theta_{10}$. Normal distributions.}
\label{tab:tcf23:1}
\begin{center}
\begin{tabular}{@{}l r r l r r r@{}}
\tblhead{
Scenario & $\theta_{20}$ & $\theta_{30}$ & $(n_1, n_2, n_3)$ & \multicolumn{3}{c}{Nominal level} \\
& & & & 0.90 & 0.95 & 0.99}
& & & (20, 20, 20) & 0.881 & 0.945 & 0.981 \\ 
& & & (30, 30, 30) & 0.881 & 0.940 & 0.990 \\ 
1 & 0.5 & 0.8 & (50, 30, 30) & 0.877 & 0.940 & 0.986 \\ 
& & & (50, 50, 50) & 0.893 & 0.947 & 0.990 \\ 
& & & (100, 50, 50) & 0.886 & 0.944 & 0.991 \\ 
& & & (100, 100, 100) & 0.883 & 0.943 & 0.988
\lastline
& & & (20, 20, 20) & 0.880 & 0.943 & 0.970 \\ 
& & & (30, 30, 30) & 0.887 & 0.950 & 0.989 \\ 
2 & 0.8 & 0.8 & (50, 30, 30) & 0.865 & 0.943 & 0.990 \\ 
& & & (50, 50, 50) & 0.878 & 0.939 & 0.989 \\ 
& & & (100, 50, 50) & 0.885 & 0.942 & 0.988 \\ 
& & & (100, 100, 100) & 0.883 & 0.943 & 0.991
\lastline
& & & (20, 20, 20) & 0.726 & 0.744 & 0.762 \\ 
& & & (30, 30, 30) & 0.852 & 0.890 & 0.909 \\ 
3 & 0.9 & 0.9 & (50, 30, 30) & 0.859 & 0.892 & 0.911 \\ 
& & & (50, 50, 50) & 0.891 & 0.947 & 0.985 \\ 
& & & (100, 50, 50) & 0.889 & 0.947 & 0.984 \\ 
& & & (100, 100, 100) & 0.879 & 0.941 & 0.989
\lastline
\end{tabular}
\end{center}
\end{footnotesize}
\end{table}

\end{document}